\newcommand{\CLASSINPUTbottomtextmargin}{4.16cm}

\documentclass[conference,a4paper]{IEEEtran}

\usepackage[left=1.43cm,right=1.43cm,top=1.8cm,bottom=4.16cm]{geometry}
\IEEEoverridecommandlockouts
\usepackage{cite}
\usepackage{amsmath,amssymb,amsfonts,bbm,siunitx}
\usepackage{graphicx}
\usepackage{textcomp}
\usepackage[dvipsnames]{xcolor}
\usepackage{multirow}
\usepackage{subcaption}
\usepackage{caption}\usepackage{tikz}
\usepackage{tkz-tab}
\usetikzlibrary{automata,arrows,positioning,calc}
\usetikzlibrary{shapes,snakes}
\usetikzlibrary{arrows}
\usepackage{dsfont}
\usepackage{soul}
\usepackage{subfiles}
\usepackage{comment}
\usepackage{algorithm}
\usepackage{algpseudocode}
\usepackage{amsmath}
\usepackage{amssymb}
\usepackage{amsthm}
\usepackage{booktabs}
\usepackage{epstopdf} 

\def\BibTeX{{\rm B\kern-.05em{\sc i\kern-.025em b}\kern-.08em
    T\kern-.1667em\lower.7ex\hbox{E}\kern-.125emX}}

\usepackage{tikz}
\usetikzlibrary{fit,calc}

\begin{document}

\bstctlcite{IEEEexample:BSTcontrol}

\title{Studying the Role of Synthetic Data for Machine Learning-based Wireless Networks Traffic Forecasting}
\author{
\IEEEauthorblockN{José Pulido$^{\mathsection, \dagger}$, Francesc Wilhelmi$^{\mathparagraph, \dagger}$\vspace{0.1cm}, Sergio Fortes$^{\mathsection, *}$,\\Alfonso Fernández-Durán$^{\ddagger}$, Lorenzo Galati Giordano$^{\dagger}$\vspace{0.1cm}, Raquel Barco$^{\mathsection}$
}
\IEEEauthorblockA{$^{\mathsection}$\emph{Universidad de Málaga, Málaga, Spain}}
\IEEEauthorblockA{$^{\dagger}$\emph{Nokia Bell Labs, Stuttgart, Germany}}
\IEEEauthorblockA{$^{\mathparagraph}$\emph{Universitat Pompeu Fabra, Barcelona, Spain}}
\IEEEauthorblockA{$^{\ddagger}$\emph{Nokia, Madrid, Spain}}
\thanks{*Corresponding author: Sergio Fortes (sergio.fortes@uma.es)}
}

\maketitle

\begin{tikzpicture}[remember picture, overlay]
\node[
    anchor=north,
    draw,
    line width=0.4pt,
    inner sep=4pt,
    text width=0.78\textwidth,
    align=center
] at ([yshift=-0.35cm]current page.north) {%
    \footnotesize
    This work has been submitted to the IEEE for possible publication.\\
    Copyright may be transferred without notice, after which this version may no longer be accessible.
};
\end{tikzpicture}
\thispagestyle{plain}
\pagestyle{plain}

\begin{abstract}
Synthetic data generation is an appealing tool for augmenting and enriching datasets, playing a crucial role in advancing artificial intelligence (AI) and machine learning (ML). Not only does synthetic data help build robust AI/ML datasets cost-effectively, but it also offers privacy-friendly solutions and bypasses the complexities of storing large data volumes. 
This paper proposes a novel method to generate synthetic data, based on first-order auto-regressive noise statistics, for large-scale Wi-Fi deployments. The approach operates with minimal real data requirements while producing statistically rich traffic patterns that effectively mimic real Access Point (AP) behavior. Experimental results show that ML models trained on our synthetic data achieve Mean Absolute Error (MAE) values within 10–15\% of those obtained using real data when evaluated on the same APs, while requiring significantly less training data. Moreover, when generalization is required, synthetic-data-trained models improve prediction accuracy by up to 50\% compared to real-data-trained baselines, thanks to the enhanced variability and diversity of the generated traces. Overall, the proposed method bridges the gap between synthetic data generation and practical Wi-Fi traffic forecasting, providing a scalable, efficient, and real-time solution for modern wireless networks.



\end{abstract}

\begin{IEEEkeywords}
Machine Learning, Traffic Forecasting, Synthetic Data Generation, Wi-Fi, 802.11
\end{IEEEkeywords}

\section{Introduction}

Artificial intelligence (AI) and machine learning (ML) have gained much prominence in communications due to promising results in areas such as network optimization, signal processing, and resource allocation~\cite{liu2019machine,ahmad2020machine, giordani2020toward}. However, AI/ML highly depend on high-quality data to be trained, validated, and evaluated. Therefore, data have become a crucial asset in modern networks, and their importance is expected to increase in the following years. Network data, while massively produced by communications systems, are often subject to strict privacy regulations, proprietary restrictions, and scalability limitations, making it difficult to collect diverse and representative datasets \cite{li2024IoT, song2021federated}. Moreover, using real-world data can be challenging due to issues such as data sparsity, imbalance, and the presence of biases~\cite{xu2018deep, yang2023bias}.

To address the shortcomings associated with real network data for training ML models, synthetic data generation emerges as a promising alternative. Synthetic data consist of artificially generated samples that mimic the statistical properties and patterns of real-world data while ensuring privacy and scalability~\cite{emam2020chap}. By using tools such as generative models, mathematical simulations, or statistical processes, one can craft comprehensive datasets that preserve key characteristics of real network data without exposing sensitive information~\cite{goncalves2020generation}. In telecommunications, synthetic data can boost next-generation ML-based solutions for tasks such as network management and optimization---e.g., traffic forecasting, anomaly detection \cite{ren2019inf}, network performance optimization \cite{bega2019deepcog}---, network planning~\cite{zhu2021network}, or AI-native radios~\cite{hoydis2020toward, bellalta2024towards}, to name a few.

\subsection{Challenges of Synthetic Data Generation}
\label{sec:challenges}

For ML models to perform well in real-world applications, the synthetic data used for their training must accurately capture the underlying phenomena, making its creation challenging. At the same time, creating synthetic data must be accomplished under reasonable cost constraints, ideally not exceeding the expense of directly acquiring the real data. Therefore, the development of effective synthetic data generation methods requires preserving a balance between realism and generalization, preventing overfitting to specific patterns while maintaining key statistical properties. Below, we identify four main challenges.

\textbf{Challenge 1 - Data resemblance (C\#1):} A central question is how much synthetic data must resemble real one, so that it remains useful for training ML models. If synthetic data misses the patterns (e.g., statistical, temporal, or structural properties) present in real data, the resulting ML models might fail to make accurate forecasts. However, if synthetic data are virtually a replica of real data, the subsequent ML models may overfit to known patterns and perform badly. Striking the right balance between fidelity and diversity is crucial. This trade-off is discussed in~\cite{ammara2025syn}, which highlights that some models may achieve high statistical fidelity but lack robustness. 

\textbf{Challenge 2 - Generalization capabilities (C\#2):} Related to the previous question, another important concern lies in the effectiveness of using synthetic data in unseen scenarios. Generative models often replicate common patterns, struggling to characterize edge situations or anomalies~\cite{feuerriegel2024generative}. This is especially critical in wireless networks, where rare events or peak traffic situations can have a significant impact~\cite{fernandes2019comprehensive}. Addressing this requires either incorporating expert domain knowledge or designing objective functions that encourage outlier exploration according to the nature of the problem, areas that are still under-explored in the literature.

\textbf{Challenge 3 - Amount of data (C\#3):} Another significant challenge is determining how much real data are needed to build robust synthetic data generators. Most generative models, especially deep learning-based approaches like Generative Adversarial Networks (GANs) or diffusion models, demand substantial volumes of high-quality real data to learn meaningful patterns~\cite{figueira2022survey}. This creates a considerable limitation when data is scarce, a common issue in wireless networks due to the high costs of collection, processing, and storage, as well as data sensitivity. While synthetic data can effectively mimic real-world distributions, the quality of the generated data is highly dependent on the volume and diversity of the original dataset~\cite{naveed2021is}. 

\textbf{Challenge 4 - Privacy (C\#4):} Finally, privacy preservation in synthetic data remains an unresolved issue. While often proposed as a privacy-preserving alternative, recent research, such as that in~\cite{zhao2025does}, demonstrates that synthetic datasets generated from real data can still leak sensitive information. Specifically, adversarial attacks like membership inference can sometimes detect if an individual's data was part of the training set, particularly when overfitting occurs. Current research lines include ensuring differential privacy or implementing post-generation audits, though these often come at the cost of utility.

\subsection{Contributions}
\label{sec:introduction}

In this paper, we explore the role of synthetic data in wireless networks and propose a novel mechanism to generate synthetic data for enriching ML models. We specifically focus on the traffic prediction problem~\cite{wilhelmi2023ai}, which we put into practice through Wi-Fi data. The main contributions of this paper are as follows:
\begin{itemize} 
\item We propose a framework for generating synthetic network data, including traffic load and number of users. Our mechanism includes an information extraction engine, from which seed knowledge is derived from real network data, and a data generation method that exploits the seed knowledge to generate new traffic patterns.
\item Using our framework, we create a synthetic dataset that expands the real Wi-Fi data from~\cite{chen2021flag}.
\item We train two different types of ML models to predict the traffic load of Wi-Fi networks using the created synthetic datasets. The subsequent evaluation, performed on real data, covers an exhaustive set of use cases to assess when and how synthetic data can be adopted in practice.
\end{itemize}

\subsection{Organization}

The remainder of this paper is organized as follows: Section~\ref{sec:related_work} reviews related work on synthetic data generation and usage in telecommunications. Section~\ref{sec:proposed_mechanism} presents the proposed framework for generating synthetic network traffic data. Section~\ref{sec:data} introduces the reference dataset used by our synthetic data generator, and then analyzes the characteristics of both real and synthetic datasets. Section~\ref{sec:performance_evaluation} evaluates the effectiveness of the derived synthetic data for training ML models for traffic prediction. Section~\ref{sec:conclusions} concludes the paper and discusses future directions.

\section{Related Work}
\label{sec:related_work}

Synthetic data generation is a well-discussed topic within the ML and data science communities, particularly for domains like images, text, and audio \cite{lu2021mach}. However, structured synthetic data---such as tabular or time series data---have also garnered significant attention, especially since the introduction of frameworks like the Synthetic Data Vault (SDV) \cite{patki2016synthetic}. SDV notably demonstrated the feasibility of using ML models to capture and replicate the statistical properties of real datasets.

In wireless networks, synthetic data serves multiple purposes, including capacity planning \cite{hussain2010capacity}, coverage prediction \cite{fauzi2022mobile}, security analyses \cite{krasic2022telecom}, and the development of new technologies \cite{wilhelmi2024s}. In this paper, we focus on generating synthetic data for ML-based network traffic forecasting. 


A classical and widely adopted approach for generating synthetic data in networking involves simulation tools. These tools emulate network protocols, user behavior and applications, and communication topologies in a virtual environment to produce controlled data with high fidelity. Simulation frameworks are especially popular in wireless and IoT domains due to their ability to recreate highly dynamic and configurable scenarios without requiring real-world deployments. If connected to the real world, simulators can become digital twins that can be updated based on their ability to obtain data from the network \cite{pulido2025digital}. Unlike generative models, simulators rely on protocol logic and traffic generation rules, providing structured and interpretable outputs.

Simulators vary widely in scope and complexity. Some target specific layers of the networking stack (e.g., packet-level simulators), while others support end-to-end application behavior, node mobility, and environment-specific parameters:
\begin{itemize}
    \item Packet generators such as Ostinato \cite{ostinato2010}, RUDE\&CRUDE \cite{rude2000}, and Seagull \cite{seagull2006}, enable manual crafting of traffic headers and packet sequences. Although valuable for protocol conformance and stress testing, their utility is limited; they are not scalable for modeling realistic, large-scale, or long-duration network traffic, and provide limited protocol support.
    \item At a higher level of abstraction, discrete-event network simulators, such as ns-3 \cite{ns3_2011} or OMNeT++ \cite{omnetpp2001}, provide full-stack modeling of network topologies, protocols, and mobility patterns. For instance, ns-3 includes Wi-Fi, LTE, and 5G modules, and has been widely used in both academic and industrial studies for realistic wireless simulations. Similarly, OMNeT++, often paired with frameworks like INET or Veins, is often used for vehicular and wireless network scenarios.
    \item More specialized network simulators have also been developed to address emerging challenges in wireless communications. For instance, Komondor \cite{wilhelmi2021komondor} is a simulator designed for prototyping novel solutions for the IEEE 802.11. It focuses on efficient MAC-layer modeling and supports advanced interference coordination and contention mechanisms, with faster execution times than full-stack tools like ns-3. This makes Komondor particularly appealing for generating exhaustive datasets to design ML solutions for particular use cases (e.g., spatial reuse optimization).
    \item Synthetic data generation approaches have recently been realized through ML models (e.g., STAN \cite{xu2020stan}) and deep generative models (e.g., PacketCGAN \cite{wang2020packetcgan}). GenAI can enable digital twins, which resemble network simulators by characterizing aspects such as network user behavior, base stations, or wireless conditions~\cite{chai2024gen}. In terms of network traffic, GenAI-based methods can automatically learn traffic patterns from real-world data and produce specified amounts of synthetic traffic~\cite{figueira2022survey}.
\end{itemize}

The use of network simulators is foundational in early testing, validation of new communication technologies, but they are also gaining attention for generating synthetic datasets to develop and benchmark ML models~\cite{wilhelmi2021usage}. For instance, in~\cite{wilhelmi2022federated}, a comprehensive dataset using network simulations was provided to train and validate federated learning models for spatial reuse optimization in decentralized wireless local area networks (WLANs). This illustrates that simulator-generated data can enable the development of distributed ML algorithms that would be challenging to develop using real-world data alone due to coordination and privacy constraints. Similarly, various studies have leveraged ns-3 simulations to generate training datasets for reinforcement learning agents in network optimization tasks, where the controlled environment allows for systematic exploration of different network conditions and policy performance evaluation~\cite{gawlowicz2019ns}.

Despite their advantages, simulation-based data generation faces several limitations. On the one hand, simulators offer full control over scenario parameters, even enabling the testing of rare edge cases, and allow for reproducible experiments with no privacy risks. On the other hand, configuring realistic traffic requires deep protocol knowledge, substantial manual effort, and typically large computational efforts. Moreover, simulated traffic often lacks the behavioral diversity observed in real-world networks, which can limit generalization when ML models trained on simulated data are applied in live deployments. Thus, while simulators remain invaluable tools in networking research, they are best used in conjunction with empirical data or for pre-training models that are later fine-tuned on real traffic.


In recent years, synthetic data generation approaches have shifted towards ML models (e.g., STAN \cite{xu2020stan}) and deep generative models (e.g., PacketCGAN \cite{wang2020packetcgan}). These methods can automatically learn traffic patterns from real-world data and produce specified amounts of synthetic traffic~\cite{figueira2022survey}. In addition, GenAI-based synthetic data are foreseen to ensure privacy by removing sensitive information.

The work in~\cite{dimyati2021time} focuses on using time-series GANs to augment telecommunications datasets, targeting traffic scenarios with temporal structure. The paper evaluates the generated data using forecasting accuracy, showing improved performance over baseline models. More recently, \cite{pandey20245gt} presented 5GT-GAN-NET, a supervised-loss GAN architecture tailored for 5G Internet traffic forecasting, where the GAN is optimized both for realism and predictive utility. The results suggest that supervised regularization stabilizes training and enhances forecasting quality.

In \cite{zhang2018gen}, the use of a GAN is proposed to produce synthetic time series datasets in smart grids sampled from the same distribution as the real datasets. To evaluate the synthetic datasets, statistical tests and classical ML tasks, such as time series clustering and load prediction, are carried out. Empirical results show that synthetic and real datasets are indistinguishable. In \cite{rotem2022transfer}, a transfer learning method for time series classification (TSC) using synthetically generated univariate time series (UTS) data is proposed. The authors proposed a synthetic time series generation algorithm, for which a convolutional neural network (CNN) model was pre-trained on the created synthetic data. The evaluation highlighted the advantages of transfer learning by reducing the training set. Finally,~\cite{chai2024gen} proposes a framework where GenAI is used to build a digital twin of the mobile network. Such a digital twin is capable of modeling users, base stations, and wireless conditions, which allows for more controlled and context-aware data synthesis, with applications in performance prediction and proactive network optimization.

Challenges with GenAI models include the risk of overfitting to training data (potentially leaking sensitive information), and a lack of generalization to rare but important events. Privacy concerns have prompted an investigation into whether synthetic data truly provides protection, as explored in~\cite{zhao2025does}. 


Our paper advances the state-of-the-art by proposing a novel, cost-efficient method for generating realistic synthetic data for wireless networks. This method directly retrieves key statistical properties from real data, such as distributions, temporal autocorrelations, and multivariate relationships, to create meaningful synthetic datasets.


\section{Proposed Synthetic Data Generation Method}
\label{sec:proposed_mechanism}

\begin{figure*}[t] 
    \centering
    \includegraphics[width=.8\textwidth]{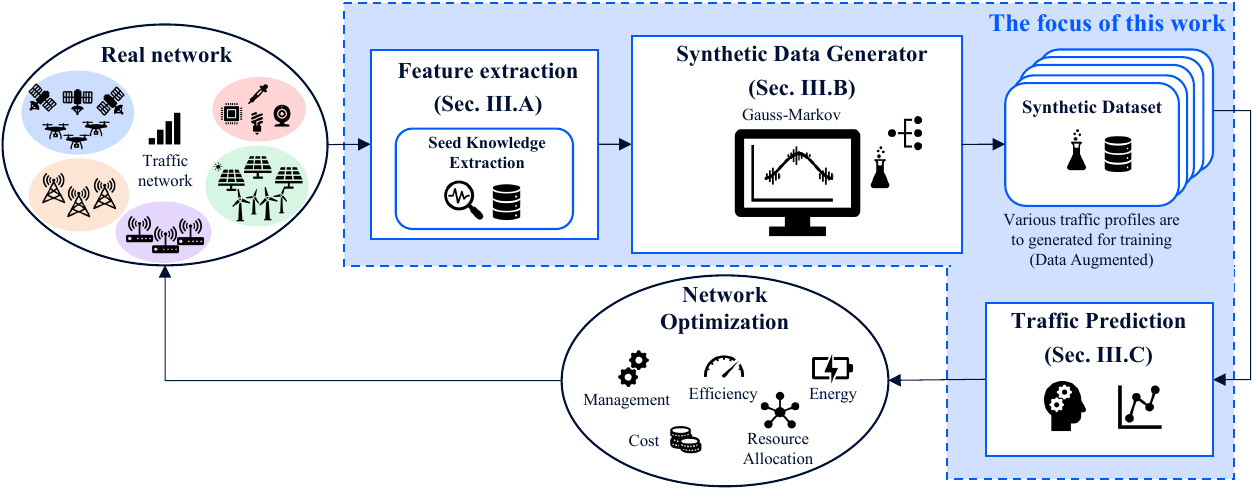}  
    \caption{Overview of ML-empowered networks with integrated synthetic data generation capabilities.}
    \label{system_overview}
\end{figure*}

This section presents the proposed solution in detail. Figure~\ref{system_overview} shows the main building blocks of the system. First, relevant network characteristics are extracted and seed knowledge is generated (e.g., traffic profiles). Using such knowledge, a synthetic data generator creates a synthetic dataset, which is then used for enriching ML models that, in this particular study, perform traffic prediction. The enhanced predictions made by the ML models can then be used to optimize the network (e.g., proactive resource allocation).

\subsection{Feature Extraction and Seed Knowledge Generation}




The foundation of our synthetic data generation approach lies in the systematic extraction of statistical characteristics from real Wi-Fi network traffic data. This process transforms raw network measurements into structured seed knowledge that captures the essential statistical properties required for realistic synthetic traffic generation.  The extracted seed knowledge serves multiple purposes in the synthetic data generation pipeline. First, it provides the baseline deterministic patterns that anchor synthetic traffic generation to realistic temporal structures. Second, it parameterizes the stochastic components that introduce controlled randomness while maintaining statistical consistency with real traffic. Third, it enables the synthetic generator to reproduce complex traffic phenomena such as peak-hour congestion, weekend usage patterns, and seasonal variations.

The feature extraction is performed on real AP time series data to identify and quantify key temporal patterns that define network behavior. The primary objective is to extract comprehensive statistical descriptors that encapsulate both the deterministic patterns and stochastic variations inherent in real network traffic. Let $L^{(k)}$ represent the AP $k$'s real traffic load measurements at discrete time instances inter-spaced by $\tau$ (set to 10 minutes to capture short-term traffic dynamics). Without loss of generality, we derive new time series $M^{(k)} = \{(\mu^{(k)}_{w,h}, \text{Var}^{(k)}_{w,h}) \mid w \in \{\text{Mon, Tue, \dots, Sun}\}, h \in \{0, 1, \dots, 23\} \}$ with statistical features extracted from the real dataset $L^{(k)}$, including the temporal mean ($\mu$) and the variance ($\text{Var}$) calculated for each hour of the day $h$ and day of the week $w$. This results in a $24 \times 7 = 168$ time series with two different values, each summarizing the essence of the traffic load behavior of AP $k$.

Starting with the aggregated mean traffic load for AP $k$ for hour $h$ and day of the week $w$, it is calculated as

\begin{equation}
    \mu^{(k)}_{w,h} = \frac{1}{|D^{(k)}_w|} \sum_{d \in D^{(k)}_w} \left( \frac{1}{N^{(k)}_{d,h}} \sum_{i=1}^{N^{(k)}_{d,h}} L^{(k)}_{d,h,i} \right) ,    
\end{equation}

where $D^{(k)}_w$ is the set of all days of week $w$ in $L^{(k)}$ and $N^{(k)}_{d,h}$ is the number of measurements within hour $h$ and day $d$ (e.g., $N^{(k)}_{d,h}=6$ for $\tau=10$ minutes). Similarly, the variance is calculated as

\begin{equation} \label{eq:compact_variance_with_mean_def}
    \text{Var}^{(k)}_{w,h} = \frac{1}{|D^{(k)}_w|} \sum_{d \in D^{(k)}_w} \left( \mu^{(k)}_{d,h} - \bar{\mu}^{(k)}_{w,h} \right)^2 ,
\end{equation}

where $\mu^{(k)}_{d,h}$ represents the mean traffic load for AP $k$ on specific day $d$ and hour $h$, and $\bar{\mu}^{(k)}_{d,h}$ is the overall mean for day of week $w$ and hour $h$ across all the observed days in the dataset, computed as

\begin{equation}
\bar{\mu}^{(k)}_{d,h} = \frac{1}{|D^{(k)}_w|} \sum_{d\in D^{(k)}_w} \mu^{(k)}_{d,h} .
\end{equation}





With the considered windowed aggregation process, which transforms raw measurements into structured and compressed statistical descriptors, it is sought to preserve essential traffic characteristics while reducing noise and measurement artifacts. Therefore, the resulting time series $M^{(k)}$ aims to preserve statistical consistency with the real network data while introducing sufficient diversity to enhance ML model training and provide generalization capabilities, as discussed in Section~\ref{sec:performance_evaluation} through empirical evaluation.

\subsection{Synthetic Data Generator}


\begin{figure}[t!]
\centering
\includegraphics[width=1.0\columnwidth]{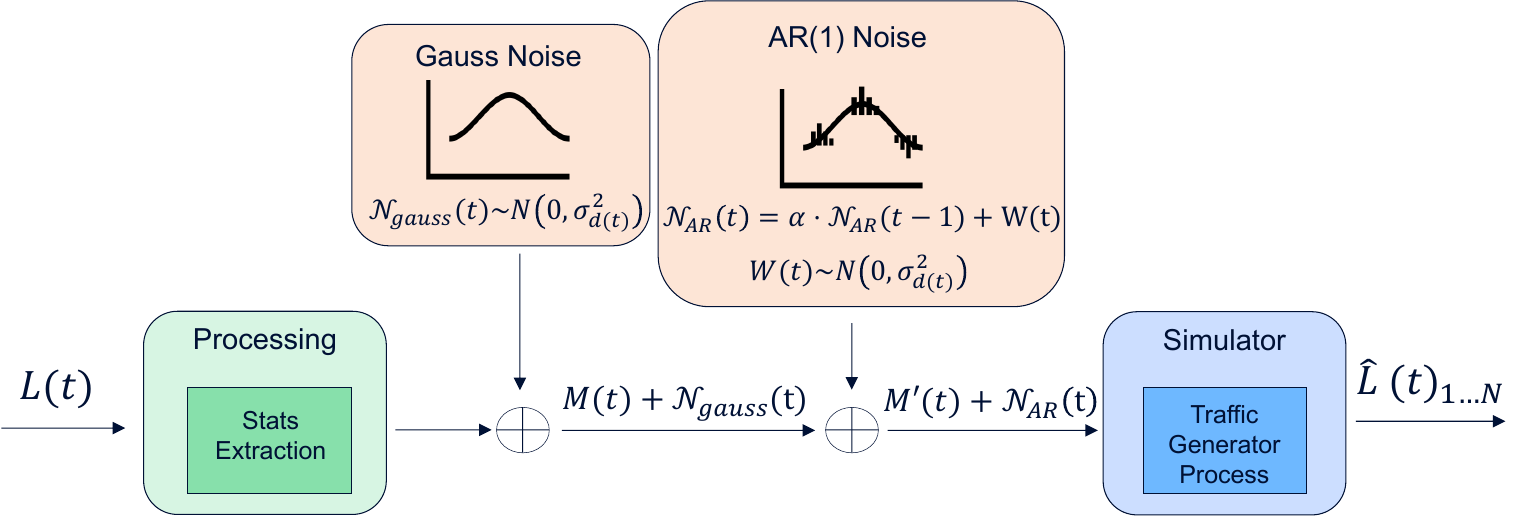}
\caption{Gauss-Markov Noise Model scheme.}
\label{gauss_markov_model}
\end{figure}



To generate synthetic data, we propose a Gauss-Markov noise-based approach---a well-established approach for modeling stochastic processes with memory---that captures both temporal dependencies and realistic traffic variations. The synthetic data generator is designed to maintain the essential properties of real-world Wi-Fi traffic patterns while introducing controlled variability. Unlike classical simulators or deep generative models, our mechanism does not aim to replicate the real time series exactly. Instead, it creates statistically grounded variations that enrich the training space of ML models. Our method builds upon the theoretical foundation of Basu \cite{basu1996time}, which generalizes the linear coding scheme of Schalkwijk and Kailath for white noise processes to autoregressive (AR) noise processes. Using an AR structure, dependencies between consecutive time steps are introduced, ensuring that the generated traffic exhibits realistic temporal dynamics.

The generator relies on seed knowledge extracted from the real AP data, including weekly patterns, hourly means, variance profiles, and short-term temporal dependencies. These descriptors summarize the combined activity of multiple users connected to the AP, capturing aggregated dynamics that are fundamental for generating meaningful synthetic sequences. Importantly, these descriptors act as statistical anchors: they preserve the large-scale structure of real traffic but do not constrain the generator to reproduce the same exact realizations. The system architecture, depicted in Fig.~\ref{gauss_markov_model}, includes the following steps: 
\begin{enumerate}
    \item Initialize an AR (Gauss-Markov) noise component, $\mathcal{N}_{\text{AR}}(0)$, for temporal correlation with the first observation from the extracted mean values.
    \item For each generated time step $t$, compute a baseline traffic pattern using a function $g(\cdot)$.
    \item Generate independent Gaussian noise $\mathcal{N}_{\text{gauss}}(t) \sim \mathcal{N}(0, \text{Var}_{d(t)})$ to apply day-of-week variability.
    \item Compute the AR noise component $\mathcal{N}_{\text{AR}}(t)$ for each generated time step $t$.
    \item Calculate the synthetic traffic data $\hat{L}(t)$ for time step $t$.
    \item Repeat steps 2-5 for the desired temporal extent (e.g., 1 week, 1 month).
\end{enumerate}

Beyond this procedural description, the generator can be interpreted through the three fundamental components that govern the statistical structure of the synthetic time series:
\begin{itemize}
\item Deterministic baseline pattern: captures the weekly and hourly structure extracted from the real AP data, ensuring that the generated sequences follow realistic large-scale temporal trends.
\item Independent Gaussian noise: introduces instantaneous fluctuations aligned with day-of-week variability, representing short-term changes in aggregated user activity.
\item Gauss–Markov autoregressive noise: injects temporally correlated variations that mimic burstiness, load persistence, and inertia effects inherent to multi-user wireless traffic.
\end{itemize}

These three components collectively introduce two complementary sources of controlled diversity:
(i) instantaneous variability driven by the Gaussian noise term, and
(ii) correlated variability driven by the AR component.
Together, they expand the space of possible traffic realizations while remaining anchored to the statistical descriptors extracted from real data. The goal is not to replicate the real time series, nor to “generalize” the data itself, but to enrich the training space in a statistically grounded manner so that downstream ML models trained on synthetic traces can achieve improved generalization performance.

The seed knowledge, consisting of the time series of mean and variance values extracted from the real AP time series $L$, along with temporal correlations, bursts, and seasonality patterns, serves as the statistical foundation for synthetic traffic generation. In particular, for a given AP $k$, synthetic time series $\hat{L}^{(k)}$ are generated as

\begin{equation}
\hat{L}^{(k)}(t) = \max(0, g(M^{(k)}(t)) + \mathcal{N}_{\text{gauss}}(t) + \mathcal{N}_{\text{AR}}(t)) .
\end{equation}




The baseline traffic pattern, aimed at realizing a cyclic traffic generation based on the weekly and hourly patterns extracted in $M$, is computed as follows:

\begin{equation}
g(t) = \mu\left[\left\lfloor \frac{t}{\tau} \right\rfloor \bmod |\mu|\right] ,
\end{equation}


where $\tau$ corresponds to the considered 10-minute sampling period, and $|\mu|$ is the length of the mean values array. This baseline serves as the anchor upon which synthetic variations are generated, ensuring that new patterns remain grounded in the statistical characteristics identified from real network behavior.







As for the AR noise component $\mathcal{N}_{\text{AR}}(t)$, which introduces temporal memory and correlation through a first-order Markov process, is calculated as

\begin{equation}
\mathcal{N}_{\text{AR}}(t) = \alpha \cdot \mathcal{N}_{\text{AR}}(t-1) + W(t) ,
\end{equation}

where $\alpha \in [0,1)$ is the AR coefficient that controls the temporal correlation strength ($\alpha \rightarrow 0$ produces nearly independent variations while $\alpha \rightarrow 1$ creates strong temporal dependencies and smoother transitions),\footnote{In our implementation, we employ $\alpha = 0.9$ to maintain strong temporal correlation while allowing sufficient stochastic variation for traffic pattern generation.} $W(t) \sim \mathcal{N}(0, \sigma^2_{d(t)})$ is independent white Gaussian noise, and $\mathcal{N}_{\text{AR}}(0)$ is initialized with the first traffic observation from the AP $k$ time series $L^{(k)}$. 

Overall, the variability injected by the independent and correlated noise components expands the space of plausible traffic behaviors while remaining rooted in the statistics of the real data. Importantly, this diversification is not meant to endow the synthetic data with any inherent “generalization” capacity. Rather, the purpose of the controlled noise is to broaden the statistical support of the generated sequences, beyond the exact realizations observed in the real traces, so that ML models trained on these synthetic datasets can ultimately achieve stronger generalization capabilities when applied to APs not seen during training.

\subsection{Network Traffic Prediction}
\label{sec:traffic_prediction}

Traffic forecasting is a crucial task in the management and optimization of wireless networks. Historically, statistical methods have been employed for traffic time series analysis and prediction. However, the increasing volume and complexity of network data have driven the adoption of ML techniques to achieve more accurate and robust predictions. 

In this context, the study presented in \cite{subhajit2023syn} addresses the prediction of daily demand for shared electric vehicles, specifically electric scooters, using ML techniques. Although focused on the demand for a mobility service, this work shares similarities with network traffic prediction in terms of time series analysis and the need to anticipate future demand.

In the following, we summarize the three main blocks involved in the prediction of network traffic using AI/ML, \textit{viz.} time series data processing, predictive function, and model performance.

\subsubsection{Time series data processing}

Network traffic prediction can be formulated as a time series forecasting problem, where the future traffic or load at a given network device or location is predicted based on previously gathered measurements. In particular, we define a time series of a given AP or BS $k\in\mathcal{K}$ as $\mathbf{X}^{(k)} \in \mathbb{R}^{N \times T}$, where $\mathbf{x}^{(k)}_t \in \mathbf{X}^{(k)}$ represents a set of features of size $N$ observed at the time step $t\in T$. To address the problem using supervised learning, we split the AP $k$'s dataset $\mathcal{D}^{(k)}$ into time series data using a sliding window of size $W$, thus $\mathbf{X}^{(k)}_t = \mathbf{X}^{(k)}[:, t:t+W], \forall t \in \mathcal{D}^{(k)}$. The windowed data are divided into features and labels, so that $\mathbf{X}^{(k)}_t = \{\mathbf{X}^{(k)}_{t-l-1:t}, \mathbf{Y}^{(k)}_{t+1:t+s}\}$, where $l$ and $s$ are referred to as lookback and number of steps.

\subsubsection{Predictive function}

A function $f(\cdot)$ is considered to predict, at a given point in time $t$, the load of a given AP $k$ based on historical observations, $\mathbf{\hat{Y}}^{(k)}_{t+1:t+s} = f(\mathbf{X}^{(k)}_{t-l-1:t})$. In particular, $f(\cdot)$ is embodied by an ANN model, which is characterized by a complex set of parameters $\theta$ (i.e., weights and biases). The values of $\theta$ can be automatically adjusted (thus trained) to the data in $\mathbf{X}^{(k)}$ by running an optimizer $o$ with a certain loss function $l(\cdot)$, so that the values predicted by $f(\cdot)$, $\mathbf{\hat{Y}}^{(k)}$, approximate their ground-truth counterpart, $\mathbf{Y}^{(k)}$. In this work, our predictive functions $f(\cdot)$ are based on CNN and LSTM architectures (refer to~\cite{wilhelmi2023ai}). 

\vspace{0.2cm}
\noindent\fbox{%
\parbox{.965\columnwidth}{%
For the remainder of this paper, models trained using real data ($\mathcal{D}_R$) are referred to as \textit{real models}, whereas models trained with synthetic data ($\mathcal{D}_S$) are called \textit{synthetic models}.
}
}
\vspace{0.2cm}




\subsubsection{Model performance}

The performance of an ML model can be quantified as a measure of prediction error, computed from the discrepancies between $\mathbf{\hat{Y}}^{(k)}$ and $\mathbf{Y}^{(k)}$. To ensure generalization capabilities, the evaluation of the model is performed in a dedicated data partition that has not been seen during the training phase. In particular, we use a training data partition $\mathbf{X}^{(k)}_\text{train} \subset \mathbf{X}^{(k)}$ to train the model, and a different test partition $\mathbf{X}^{(k)}_\text{test} \subset \mathbf{X}^{(k)}$ to evaluate the performance of the model, which meet the requirement $\mathbf{X}^{(k)}_\text{train} \cap \mathbf{X}^{(k)}_\text{test}$. In this paper, we specifically adopt the Mean Absolute Error (MAE) as the main evaluation metric. The MAE of the predictions in a given AP $k$ is computed as 
\begin{equation}
    \text{MAE}(k) = \frac{1}{|\mathbf{X}^{(k)}_\text{test}|} \sum_{\mathbf{y}_t \in \mathbf{X}^{(k)}_\text{test}} |\mathbf{y}_t -\mathbf{\hat{y}}_t| .
\end{equation}



\section{Synthetic Data Analysis and Validation}
\label{sec:data}

\begin{figure}[t!]
    \centering
    \begin{subfigure}{0.7\columnwidth}
        \centering
        \includegraphics[width=\linewidth]{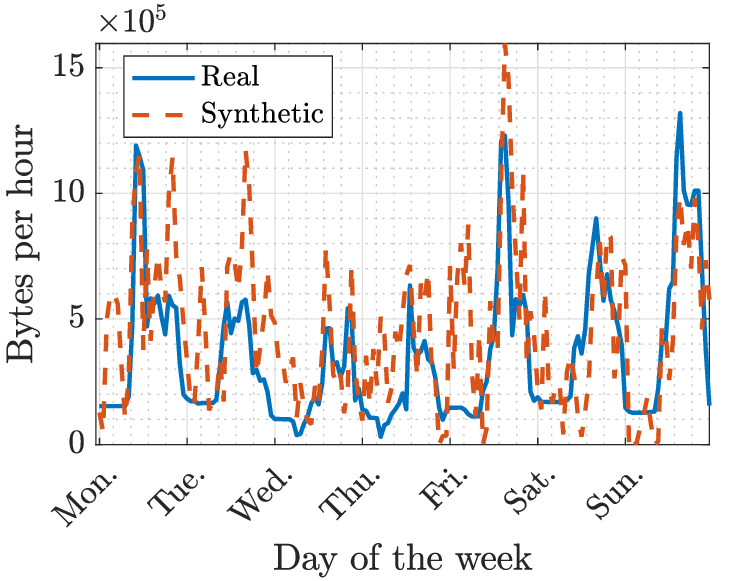}
        \caption{Mean per hour}
        \label{fig:bytes_tx_week_per_hour}
    \end{subfigure}
    \hfill
    \begin{subfigure}{0.7\columnwidth}
        \centering
        \includegraphics[width=\linewidth]{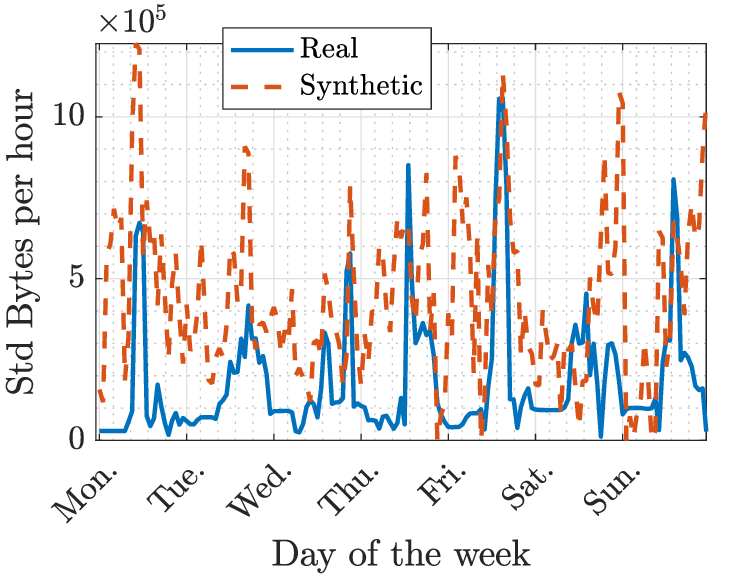}
        \caption{Standard deviation per hour}
        \label{fig:std_week_per_hour}
    \end{subfigure}
    \caption{Mean and standard deviation of the Bytes transmitted over one week by a real AP and its synthetic counterpart.}
    \label{fig:data_analysis_week}
\end{figure}

\subsection{Considered Dataset}
\label{sec:dataset_presentation}

The experiments carried out in this paper are based on the open-source dataset from~\cite{chen2021flag}, which contains measurements collected from 7404 Wi-Fi APs during $49$ days. As done in~\cite{wilhelmi2023ai}, we select a random subset of 100 APs and derive the load of each of them after downsampling it in windows of $10$ minutes. The load of each considered AP is transformed along with other available features (number of connected users, hour of the day, and day of the week) into time series arrays, as previously indicated in Section~\ref{sec:traffic_prediction}. In particular, the selected features correspond to \textit{i)} total (uplink and downlink) AP traffic load, \textit{ii)} average number of connected (active) users, \textit{iii)} hour of the day (transformed into sine/cosine values), and \textit{iv)} day of the week (transformed into sine/cosine values).

In the subsequent stage of the analysis, the dataset will be examined in conjunction with the synthetic data generated using the Gauss-Markov model described in Section~\ref{sec:proposed_mechanism}.

\subsection{Case Study: Analysis of a Single Access Point}
\label{sec:dataset_analysis}

A thorough examination of a single exemplar AP (i.e., AP 7-1389 in the dataset~\cite{chen2021flag}) is conducted to assess and compare the behavior of the real Wi-Fi network's traffic (containing its traffic over 15 days) and the synthetic counterpart. The aim of our synthetic generation method is to generate new samples that maintain essential statistical patterns and structures, rather than creating exact replicas of the real data, which would not add value to ML model training. 

To showcase the relationship between real and synthetically generated data, Fig.~\ref{fig:data_analysis_week} depicts the temporary traffic of our sample AP for the first week of measurements, both in terms of mean transmitted bytes and their standard deviation. As shown in Fig.~\ref{fig:bytes_tx_week_per_hour}, the average number of bytes generated by our synthetic generator closely resembles those from the real AP. However, as shown in Fig.~\ref{fig:std_week_per_hour}, the standard deviation of synthetic data is greater than for real data due to the added noise, thus providing the necessary variability to improve the ML models.

\begin{figure}[t!]
    \centering
    \includegraphics[width=.95\linewidth]{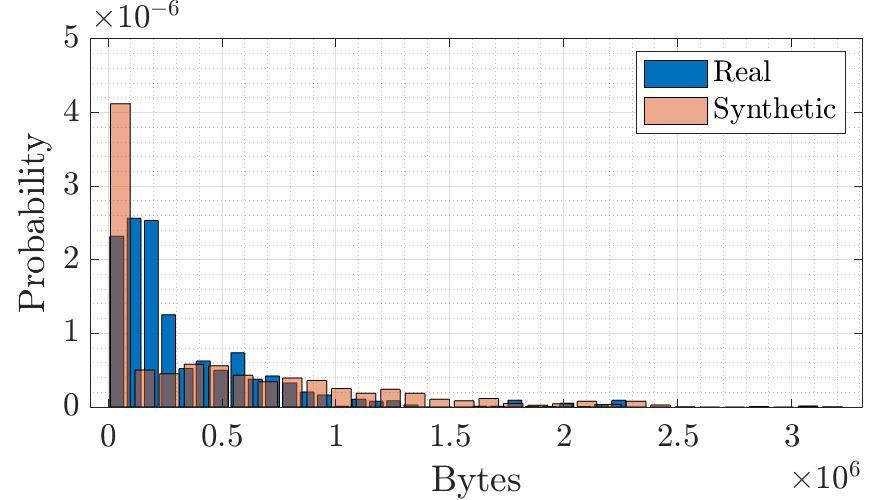}
    \caption{Distribution of the bytes transmitted by a real AP and its synthetic counterpart.}
    \label{distribution_tx_bytes}
\end{figure}

To further highlight the differences between real and synthetic traces, Fig.~\ref{distribution_tx_bytes} shows the histograms of the load for each approach. While the two different datasets do not conform to the same distribution, there are certain ranges where they are similar, such as between 0.5 and 1 MB.

\begin{figure}[t!]
\centering
\includegraphics[width=0.85\columnwidth]{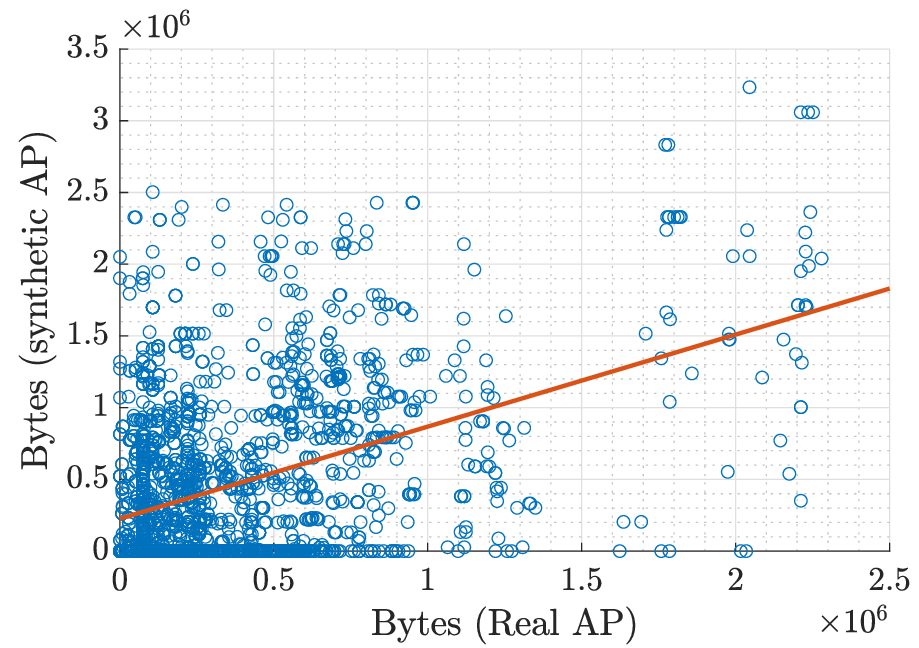}
\caption{Correlation between the real and synthetic data.}
\label{fig:correlation}
\end{figure}

Next, Fig.~\ref{fig:correlation} illustrates a correlation plot between the average transmitted bytes of the real AP data and its synthetic counterpart. Each point corresponds to a ``transmitted bytes'' data instance. The red line represents the best linear fit, showing the degree of linear association between the two datasets. The positive trend with moderate dispersion indicates that, although not perfectly aligned, the synthetic generation process captures the general behavior of average traffic volumes. In turn, the shown disparity among data points exemplifies, once again, the statistical variations between real and synthetic datasets.

\begin{figure}[ht]
    \centering
    \begin{subfigure}{0.8\columnwidth}
        \centering
        \includegraphics[width=\linewidth]{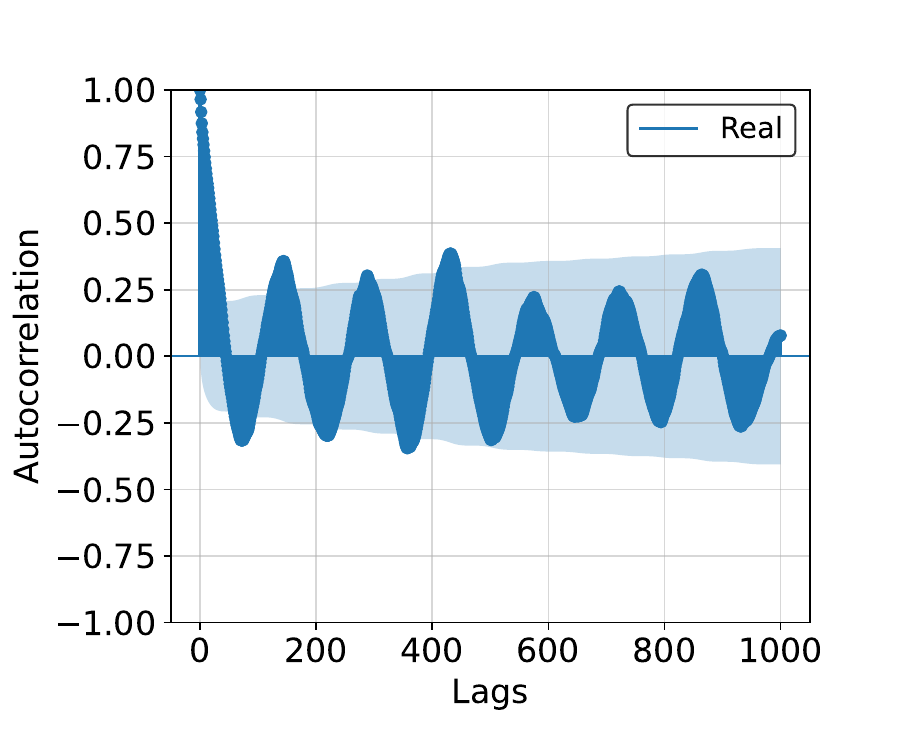}
        \caption{Real data}
        \label{fig:autocorrelation_real}
    \end{subfigure}
    \hfill
    \begin{subfigure}{0.8\columnwidth}
        \centering
        \includegraphics[width=\linewidth]{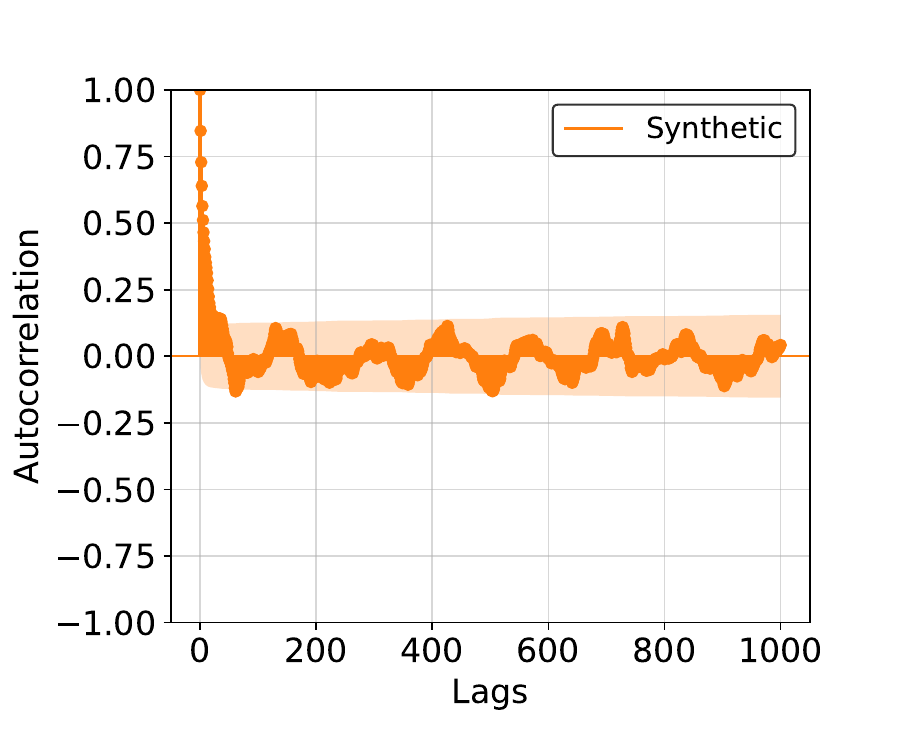}
        \caption{Synthetic data}        
        \label{fig:autocorrelation_synthetic}
    \end{subfigure}
    \caption{Autocorrelation of the data in the real AP (left) and its synthetic counterpart (right).}
    \label{fig:autocorrelations}
\end{figure}

Finally, Fig.~\ref{fig:autocorrelations} presents the autocorrelation plots of the real and synthetic AP traffic data, computed at different lags, i.e., at different time steps in the future (with lag 1 and lag 6 corresponding to 10 minutes and 1 hour, respectively). These lags were selected to capture short-term and mid-term temporal dependencies in the data. The autocorrelation at lag 1 provides insights into the immediate continuity and smoothness of the traffic behavior, while lag 6 captures repetitive or periodic patterns at an hourly scale, potentially reflecting daily user behavior or device connectivity cycles. On the real data (Fig.~\ref{fig:autocorrelation_real}), we observe a strong autocorrelation at both lags, suggesting the presence of regular traffic patterns and short-term temporal dependencies, which are common in Wi-Fi usage due to user routines and background data transmissions. The synthetic data (Fig.~\ref{fig:autocorrelation_synthetic}) exhibits similar autocorrelation behavior, although with slightly reduced magnitude, indicating that the generative process is able to replicate the temporal structure to a reasonable extent. These results suggest that the synthetic data not only replicates distributional properties but also retains key seasonal patterns and temporal dynamics. This is a crucial aspect for downstream tasks such as traffic prediction, where the temporal consistency of the data plays a vital role in model performance.

\subsection{Case Study: Analysis Across Multiple APs}


Here, we perform a broader statistical comparison across multiple APs to evaluate the generalizability and robustness of the generated synthetic data. This comparison involves computing key statistical metrics on both the real and synthetic datasets for several APs, and analyzing the absolute differences between them. The considered metrics cover aspects of data distribution (mean, standard deviation, coefficient of variation), temporal behavior and trends (skewness, kurtosis, autocorrelation with lag 1 and 6), and correlation of weekly traffic patterns (Pearson correlation of mean and standard deviation of bytes transmitted per day of the week). This multi-faceted analysis enables a quantitative understanding of how well the synthetic data preserves critical characteristics of real-world network traffic across various APs.

Table~\ref{table:stat_deltas} presents the statistical comparison across multiple APs (the first 20 APs in the dataset, for the sake of space), organized into four main categories: distribution metrics, higher-order moments, temporal characteristics, and weekly correlations. The results reveal several important insights regarding the fidelity of the synthetic data, which are detailed next.

\textbf{Distribution Metrics Analysis:} The differences between mean values ($\Delta\mu$) show considerable variation across APs, ranging from 3.9~KB (AP 7-129) to 424.3~KB (AP 7-323), with an overall average of 110.6 KB and standard deviation of 118.9~KB. This indicates that, while some APs exhibit excellent mean preservation, others show discrepancies. Similarly, the standard deviation differences ($\Delta\sigma$) range from 5.1~KB to 682.3~KB, suggesting that the Gauss-Markov model captures traffic variability differently across various AP profiles (from almost empty to heavily loaded APs). The coefficient of variation differences ($\Delta C_v$) present variations from 0.3\% to 882.3\%, indicating that relative variability preservation is particularly challenging for APs with specific traffic characteristics.

\textbf{Higher-Order Moments Analysis:} The preservation of skewness ($\Delta\gamma_1$) and kurtosis ($\Delta\gamma_2$) reveals systematic challenges in replicating the shape characteristics of traffic distributions. Skewness differences average 4.01 with a standard deviation of 3.93, while kurtosis differences show even greater variability (mean = 87.40, std = 132.70). Notable outliers include AP 7-111 with $\Delta\gamma_2$ = 471.68, suggesting that APs with highly asymmetric or heavy-tailed distributions are particularly difficult to model accurately. 

\textbf{Temporal Characteristics Analysis:} Autocorrelation preservation shows more consistent performance across APs. The lag-1 autocorrelation differences ($\Delta AC_1$) average 0.21 with a standard deviation of 0.09, indicating relatively stable short-term temporal dependency preservation. However, lag-6 autocorrelation differences ($\Delta AC_6$) show greater variability (mean = 0.31, std = 0.15), suggesting that longer-term temporal patterns are more challenging to replicate. The smaller standard deviation for $\Delta AC_1$ compared to $\Delta AC_6$ confirms that immediate temporal dependencies are better preserved than medium-term patterns.

\textbf{Weekly Pattern Correlations:} The weekly correlation analysis demonstrates one of the strongest aspects of the synthetic data generation. Pearson correlations for weekly means ($\rho_{p,\mu}$) average 0.84 with a relatively small standard deviation of 0.14, indicating robust preservation of weekly traffic patterns across most APs. Similarly, correlations for weekly standard deviations ($\rho_{p,\sigma}$) average 0.83 with a standard deviation of 0.22. Notably, 15 out of 20 APs (75.0\%) achieve correlations above 0.8 for weekly means, demonstrating the model's capability to capture cyclical traffic behaviors.
Apart from that, we observe that certain synthetic APs perform consistently across multiple metrics. For instance, AP 7-115 achieves excellent weekly correlations ($\rho_{p,\mu}$ = 0.99, $\rho_{p,\sigma}$ = 0.96) while maintaining moderate distribution differences. Conversely, AP 7-323 shows poor performance in the weekly standard deviation correlation ($\rho_{p,\sigma}$ = 0.05) but reasonable performance in other metrics.

\subsection{Findings on the Generated Synthetic Data}

Below, we link the main findings from the analysis conducted in this section with respect to the challenges enumerated in Section~\ref{sec:challenges}.

Starting with challenge \textit{C\#1 (Data resemblance)}, the conducted analysis reveals that the Gauss-Markov approach excels at preserving temporal structures and weekly patterns—critical characteristics for ML model training—while struggling with extreme distributional properties and higher-order moments. The consistent high weekly correlations (average $>0.8$) indicate that the synthetic data maintains the essential cyclical behaviors needed for traffic prediction tasks. However, the high variability in the coefficient of variation and kurtosis differences suggests that models requiring precise distributional fidelity may experience performance limitations.

Regarding challenge \textit{C\#2 (Generalization capabilities)}, our findings indicate that the Gauss-Markov model achieves a good balance between fidelity and diversity. While perfect statistical replication is not achieved---as evidenced by the considerable variations in higher-order moments---this controlled divergence may be beneficial for model robustness. As highlighted in recent work by Ammara et al.~\cite{ammara2025syn}, models that achieve high statistical fidelity may lack the variability needed for robustness against unseen scenarios. In our case, the preserved temporal dependencies ($\Delta AC_1$ = 0.21 ± 0.09), combined with the introduced variability in distributional characteristics, may provide a strong foundation for training generalizable traffic prediction models. Beyond that, the observed divergence in higher-order moments (mean $\Delta\gamma_2$ = 87.40) can potentially help prevent overfitting. This is especially important for wireless networks, where traffic patterns can vary greatly due to environmental factors, user behavior, or changes in network setup. By maintaining core temporal structures while introducing controlled variability, our synthetic data align with the principle that effective synthetic data should preserve essential characteristics and, at the same time, provide enough diversity to improve model generalization.

As for challenge \textit{C\#3 (Amount of data)}, our analysis across 20 APs demonstrates that the Gauss-Markov approach can generate useful synthetic data even with limited real data samples, addressing concerns raised by Naveed et al.~\cite{naveed2021is} about the substantial data requirements of deep learning-based generative models. The consistent performance in temporal characteristics preservation across diverse AP profiles suggests that the model can learn meaningful patterns without requiring extensive datasets, making it particularly suitable for privacy-sensitive wireless networking environments where labeled data is scarce or expensive to collect.

Finally, regarding challenge \textit{C\#4 (Privacy)}, our statistical approach extracts only aggregate temporal and variance characteristics, which contribute significantly to privacy preservation. In particular, our feature extraction process computes means and variances over spaced time windows, which reduces the granularity of the real data and allows anonymizing individual user activities. Ensuring privacy is critical for enabling unrestricted data sharing for ML, which would definitively foster collaboration and accelerate innovation in the area.

\begin{table*}[ht!]
\caption{Statistical analysis for 20 APs, including distribution differences ($\Delta$), temporal characteristics differences, and weekly traffic correlations between real and synthetic data.}
\label{table:stat_deltas}
\footnotesize
\centering
\begin{tabular}{l|ccc|cc|cc|cc}
\toprule
\textbf{AP ID} & \multicolumn{3}{c|}{\textbf{Distribution Metrics}} & \multicolumn{2}{c|}{\textbf{Higher-Order}} & \multicolumn{2}{c|}{\textbf{Temporal}} & \multicolumn{2}{c}{\textbf{Weekly Correlations}} \\
\cmidrule(r){2-4} \cmidrule(r){5-6} \cmidrule(r){7-8} \cmidrule(r){9-10}
 & $\Delta\mu$ (KB) & $\Delta\sigma$ (KB) & $\Delta C_v$ (\%) & $\Delta\gamma_1$ & $\Delta\gamma_2$ & $\Delta AC_1$ & $\Delta AC_6$ & $\rho_{p,\mu}$ & $\rho_{p,\sigma}$ \\
\midrule

7-41   & 174.8 & 254.0 & 0.3 & 0.42 & 3.55 & 0.26 & 0.31 & 0.82 & 0.95 \\
7-107  & 52.9 & 73.0 & 129.0 & 3.46 & 52.14 & 0.24 & 0.09 & 0.97 & 0.96 \\
7-108  & 10.7 & 14.1 & 471.0 & 5.94 & 83.83 & 0.15 & 0.05 & 0.96 & 0.98 \\
7-109  & 6.4 & 6.8 & 334.7 & 5.85 & 93.84 & 0.16 & 0.17 & 0.88 & 0.90 \\
7-110  & 6.9 & 8.9 & 176.1 & 4.37 & 57.16 & 0.20 & 0.15 & 0.94 & 0.97 \\
7-111  & 72.8 & 86.7 & 128.0 & 10.63 & 471.68 & 0.09 & 0.17 & 0.83 & 0.81 \\
7-112  & 8.7 & 8.1 & 296.9 & 3.92 & 40.69 & 0.07 & 0.14 & 0.72 & 0.82 \\
7-115  & 23.7 & 21.3 & 249.1 & 5.86 & 113.47 & 0.04 & 0.16 & 0.99 & 0.96 \\
7-117  & 248.3 & 316.8 & 18.6 & 0.01 & 4.90 & 0.32 & 0.47 & 0.75 & 0.20 \\
7-125  & 46.9 & 117.5 & 882.3 & 11.58 & 269.39 & 0.13 & 0.34 & 0.97 & 0.85 \\
7-126  & 6.4 & 5.5 & 529.8 & 10.64 & 214.60 & 0.10 & 0.29 & 0.83 & 0.98 \\
7-129  & 3.9 & 5.1 & 627.5 & 11.01 & 240.10 & 0.11 & 0.43 & 0.98 & 0.97 \\
7-131  & 102.5 & 134.0 & 20.2 & 2.27 & 24.05 & 0.22 & 0.12 & 0.88 & 0.90 \\
7-136  & 92.0 & 215.3 & 31.0 & 0.14 & 0.36 & 0.30 & 0.52 & 0.64 & 0.92 \\
7-276  & 284.4 & 367.8 & 4.9 & 0.13 & 0.26 & 0.34 & 0.53 & 0.65 & 0.90 \\
7-323  & 424.3 & 682.3 & 25.9 & 0.55 & 1.10 & 0.22 & 0.41 & 0.74 & 0.05 \\
7-555  & 159.7 & 352.8 & 33.1 & 0.82 & 2.75 & 0.28 & 0.47 & 0.96 & 0.94 \\
7-590  & 109.3 & 198.0 & 17.4 & 1.04 & 6.21 & 0.25 & 0.38 & 0.55 & 0.88 \\
7-827  & 198.9 & 228.2 & 50.76 & 1.15 & 7.73 & 0.33 & 0.43 & 0.94 & 0.89 \\
7-1389 & 188.5 & 259.6 & 12.2 & 0.28 & 1.36 & 0.29 & 0.45 & 0.57 & 0.88 \\
\midrule
\textbf{Mean} & \textbf{110.6} & \textbf{178.5} & \textbf{201.5} & \textbf{4.01} & \textbf{87.40} & \textbf{0.21} & \textbf{0.31} & \textbf{0.84} & \textbf{0.83} \\
\textbf{Std} & \textbf{118.9} & \textbf{187.3} & \textbf{267.6} & \textbf{3.93} & \textbf{132.70} & \textbf{0.09} & \textbf{0.15} & \textbf{0.14} & \textbf{0.22} \\
\bottomrule
\end{tabular}
\begin{flushleft}
\footnotesize
\textbf{Legend:} $\Delta\mu$: Mean difference; $\Delta\sigma$: Standard deviation difference; $\Delta C_v$: Coefficient of variation difference; $\Delta\gamma_1$: Skewness difference; $\Delta\gamma_2$: Kurtosis difference; $\Delta AC_1$: Lag-1 autocorrelation difference; $\Delta AC_6$: Lag-6 autocorrelation difference; $\rho_{p,\mu}$: Pearson correlation of weekly means; $\rho_{p,\sigma}$: Pearson correlation of weekly standard deviations.
\end{flushleft}
\end{table*}

These findings demonstrate that even if the synthetic data generator does not achieve perfect statistical replication, it succeeds in capturing the most relevant characteristics for network traffic modeling, i.e., temporal dependencies and cyclical patterns. The preserved weekly correlations and reasonable autocorrelation differences indicate that the synthetic data maintains sufficient structural integrity for training robust ML models. In addition, the controlled divergence in higher-order moments provides the beneficial diversity needed to prevent overfitting and enhance model generalization to unseen traffic scenarios. 

In the following, we will further validate our findings by assessing the performance of ML models trained using synthetic data.

\section{Prediction Performance Validation}
\label{sec:performance_evaluation}






\subsection{Experiments Description}

A design of experiments (DoE) has been done to analyze the impact of various experimental setups on the ability of synthetic ML models to predict the traffic of given APs. This design seeks to answer key questions about the usefulness of synthetic data and the favorable conditions for its generation and use in model training. We consider the following main degrees of freedom for the experiments (see Table~\ref{tab:design_experiments}):
\begin{enumerate}
    \item \textbf{Length of seed data}: The extension of the real data measurements (in number of days) considered to generate synthetic data according to the method introduced in Section~\ref{sec:proposed_mechanism}.
    \item \textbf{Length of train/test data}: The extension of the data (in number of days) considered for training and evaluating both real and synthetic datasets.
    \item \textbf{Number of train/test APs}: The number of APs (real or synthetic) considered for extracting the data measurements used to train and evaluate the ML models. Depending on the experiment, the APs used to train and test the ML models are the same (partitions are clearly split), but in some other cases, different sets of APs are used.
    \item \textbf{Number of experiments}: The number of repetitions considered for averaging purposes. In this case, each experiment leads to generating and evaluating an ML model based on the data from randomly selected APs from the whole dataset.
\end{enumerate}

This factorial design allows us to systematically explore interactions between selected factors, such as the combined impact of the generation approach and the number of APs, or how the temporal extent of the data affects the accuracy of the model. The inclusion of multiple experiments ensures a statistically valid assessment, while the variation in the number of APs reflects real scenarios of Wi-Fi deployments with different scales. With this approach, not only validating the effectiveness of the synthetic data is sought, but also identifying optimal policies for use in prediction tasks, providing a practical guide for future studies and applications in Wi-Fi networks.

\begin{table}[t!]
\caption{Parameters considered for each set of experiments.}
\label{tab:design_experiments}
\resizebox{\columnwidth}{!}{%
\begin{tabular}{@{}lccccc@{}}
\toprule
 & \begin{tabular}[c]{@{}c@{}}Length \\ of seed data\end{tabular} & \begin{tabular}[c]{@{}c@{}}Length \\ of train/test data\end{tabular} & \# Train AP & \# Test AP & \# Exp. \\ \midrule
Sec.~\ref{sec:exp1} & \{1, 7, 15\} days & 30 days & 1 & 1 & 100 \\
Sec.~\ref{sec:exp2} & 15 days & \{5, 15, 30, 60\} days & \{10, 20, 50\} & \{10, 20, 50\} & 10 \\
Sec.~\ref{sec:exp3} & 15 days & 60 days & \{10, 20, 50\} & \{10, 20, 50\} & 10 \\ \bottomrule
\end{tabular}%
}
\end{table}


\subsection{How much real data is needed to generate valid synthetic data?}
\label{sec:exp1}

\begin{figure}[t!]
    \centering
    \includegraphics[width=.8\linewidth]{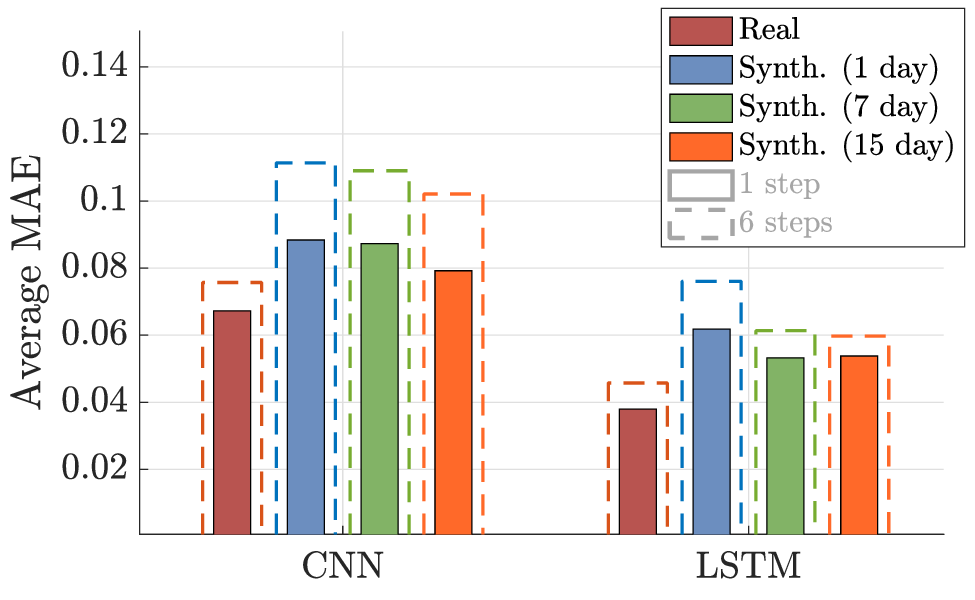}
    \caption{\textit{How much real data is needed to generate valid synthetic data?} Predictive performance (average MAE) achieved by each model ($m\in\{\texttt{CNN}, \texttt{LSTM}\}$) at different prediction horizons ($s \in\{1, 6\}$ steps) in $K=100$ APs when trained with synthetic datasets $\mathcal{D}^{(k)}_S$ generated from $|\mathcal{D}^{(k)}_R| \in \{1, 7, 15\}$ days of real measurements. The \textit{Real} baseline uses $|\mathcal{D}^{(k)}_R| = 30$ days of data.}
    \label{fig:how_much_real_data_is_needed}
\end{figure}

We first examine the requirements, in terms of seed data from real measurements, to generate compelling synthetic data, as described in Section~\ref{sec:proposed_mechanism}. For that, we consider individual ML models, that is, models individually trained and evaluated on the separate data partitions (train and test) of single APs. Those models are trained using synthetic datasets $\mathcal{D}^{(k)}_S$ of size $|\mathcal{D}^{(k)}_S| = 30$ days, which are generated from increasingly seed data extracted from the real measurements, $|\mathcal{D}^{(k)}_R| \in \{1, 7, 15\}$ days. Using this setup, Fig.~\ref{fig:how_much_real_data_is_needed} shows the average MAE obtained by the two considered ML models ($m\in\{\texttt{CNN}, \texttt{LSTM}\}$) at $s\in\{1,6\}$ across $N=100$ experiments. The baseline results obtained for $|\mathcal{D}^{(k)}_R| = 30$ days are also included.

As shown in Fig.~\ref{fig:how_much_real_data_is_needed}, there is a clear trend among the synthetically trained models toward decreasing their average MAE as the seed data size increases. This occurs for both ML models (CNN and LSTM) and for the two different prediction targets ($s=1$ and $s=6$ steps), which confirms that the more seed data available for generating synthetic data, the better the performance of the synthetically trained ML models.

However, we also observe that the synthetic models cannot beat the real ones and that the accuracy improvement is not significant even if more seed data is used, which is due to the limitations of the training performed on an individual basis. This motivates us to look at ML models that aggregate data from multiple AP sources in the following subsections.

\subsection{How much synthetic data is needed?}
\label{sec:exp2}


\begin{figure*}[t!]
    \centering
    \begin{subfigure}{0.85\columnwidth}
        \centering
        \includegraphics[width=\linewidth]{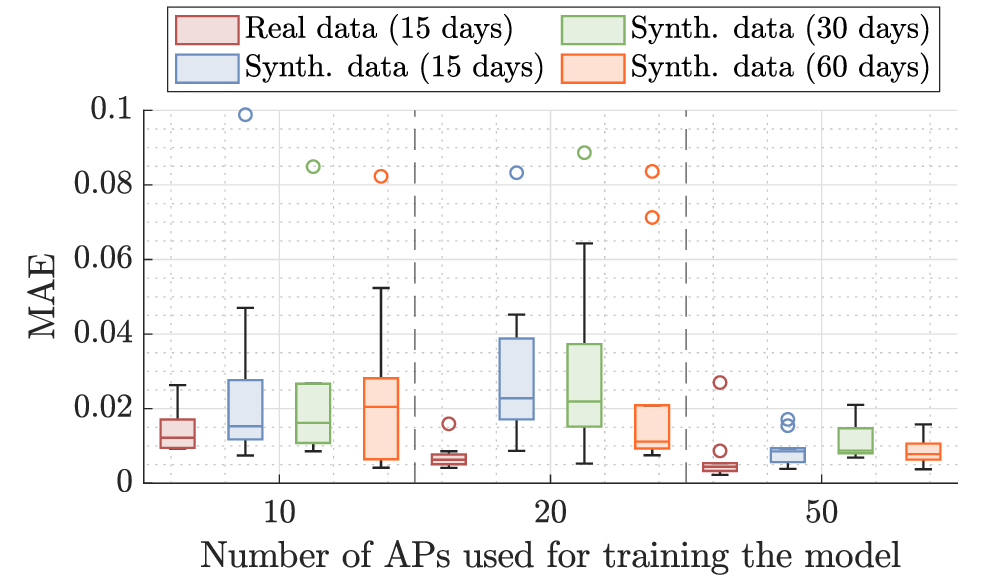}
        \caption{CNN (1 step)}
        \label{subfig:how_much_data_is_needed_a}
    \end{subfigure}
    \begin{subfigure}{0.85\columnwidth}
        \centering
        \includegraphics[width=\linewidth]{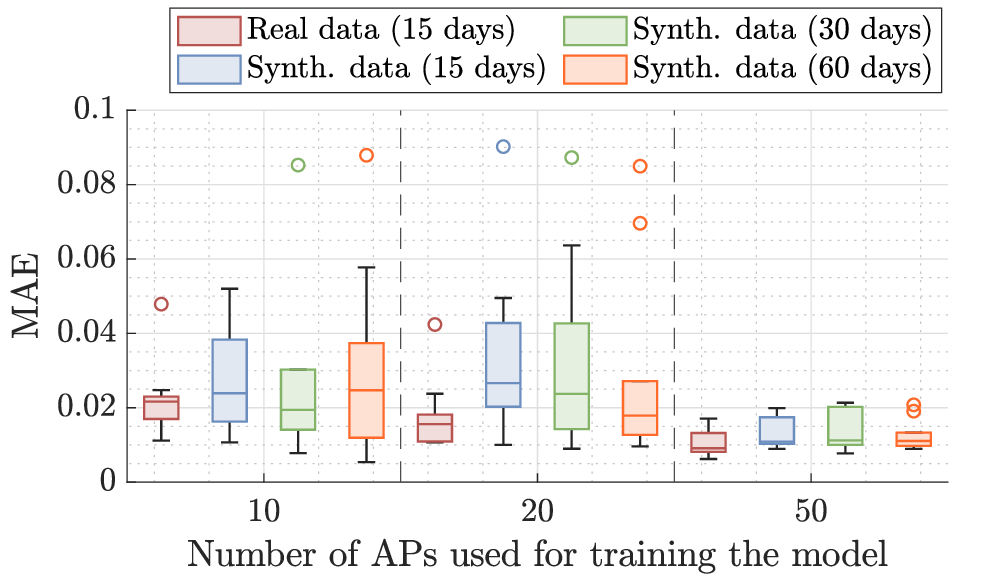}
        \caption{CNN (6 steps)}
        \label{subfig:how_much_data_is_needed_b}
    \end{subfigure}
    \begin{subfigure}{0.85\columnwidth}
        \centering
        \includegraphics[width=\linewidth]{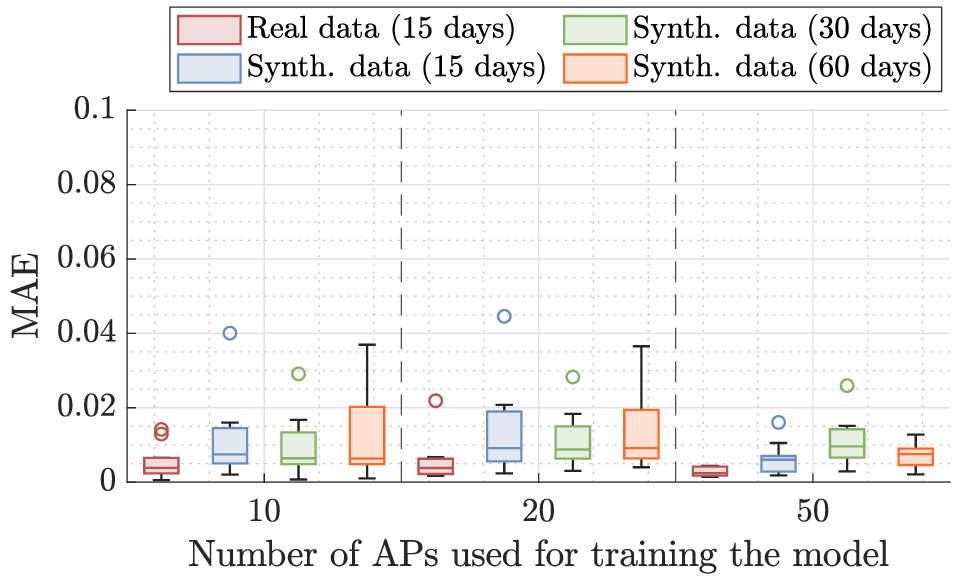}
        \caption{LSTM (1 step)}
        \label{subfig:how_much_data_is_needed_c}
    \end{subfigure}
    \begin{subfigure}{0.85\columnwidth}
        \centering
        \includegraphics[width=\linewidth]{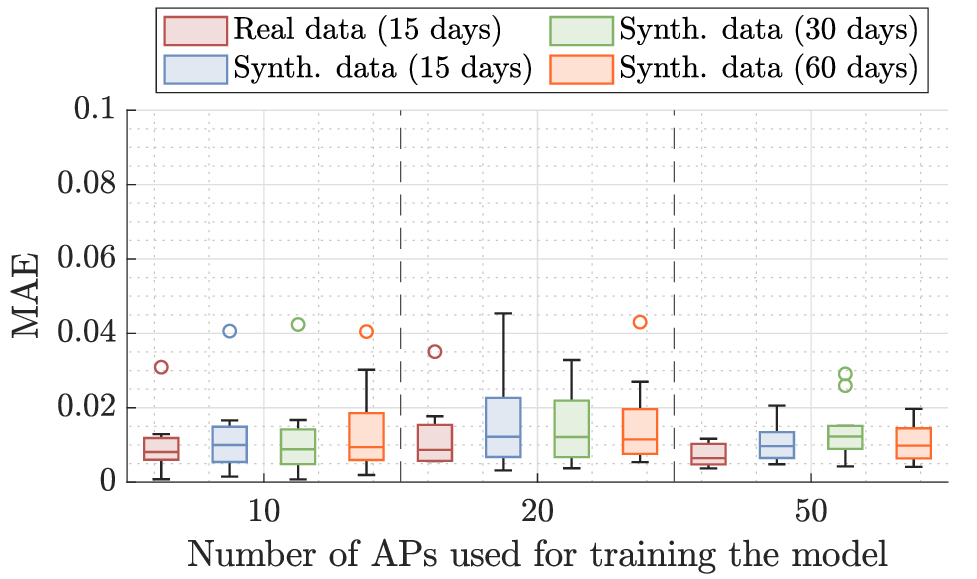}
        \caption{LSTM (6 steps)}
        \label{subfig:how_much_data_is_needed_d}
    \end{subfigure}%
    \caption{\textit{How much synthetic data is needed?} Predictive performance (MAE) achieved by each model ($m\in\{\texttt{CNN}, \texttt{LSTM}\}$) at different prediction horizons ($s \in\{1, 6\}$ steps) when trained with different amounts of real and synthetic data.}
    \label{fig:how_much_data_is_needed}
\end{figure*}

We now analyze the impact of the length of synthetic datasets on the performance of the ML models studied. For that, we look at two distinct aspects, namely \emph{i)} the number of APs from which data is used to train the models ($K\in\{10,20,50\}$ APs) and \emph{ii)} the length of the measurements taken per AP ($|\mathcal{D}^{(k)}_S|\in\{15, 30,  60\}$ days). The results are shown in Fig.~\ref{fig:how_much_data_is_needed}, where each subplot shows the predictive error (MAE) of each model, $m\in\{\texttt{CNN}, \texttt{LSTM}\}$, in two different prediction steps, $s\in\{1,6\}$, as a function of the number of APs (x-axis) and the length of their datasets (boxplots\footnote{Each boxplot includes the median (line inside the box), the worst/best 25\% performance quartiles (top and bottom edges of the box, respectively), and
the maximum/minimum values (whisker lines outside the box).} colored differently). The results obtained using real data ($\mathcal{D}_R$) are also included for each number of APs, $K$. However, due to the limitations in terms of real data measurements extension (which is precisely the root problem motivating this paper), we only consider that the length of the real datasets is limited to 15 days, i.e., $|\mathcal{D}^{(k)}_R| = 15$ days.

Starting with the results obtained by the CNN models (Fig.~\ref{subfig:how_much_data_is_needed_a} and Fig.~\ref{subfig:how_much_data_is_needed_b}), we observe that the trends are very similar for both predicted steps ($s\in\{1,6\}$). This is a first appealing property from the synthetic models, whose behavior is consistent with the one observed from the real models and confirms that the synthetic data generation process is faithful to the trends and seasonality from the real data. For a small number of APs ($K=10$ and $K=20$), the synthetic models lead to a significantly higher prediction error than the real models, especially for the smallest dataset lengths ($|\mathcal{D}_S| = \{15,30\}$ days). In some particular cases, having access to more synthetic data ($|\mathcal{D}_S| = \{60\}$ days), even from a limited set of APs ($K=10$) allows achieving better performance, as shown by the lowest 25th-quartile and the minimum values. Then, as $K$ increases, we observe that the performance of the ML models become more stable. As a result, for $K=50$ APs, the synthetic models end up being equivalent to the real ones (they lead to similar performance) and might even offer better accuracy in some cases ($s=6$ steps and $|\mathcal{D}_S| = 60$ days). In this case, a large amount of synthetic data is helpful in improving the performance of CNNs, which typically require a lot of data to perform well, at the most challenging setup ($K=50$ APs and $s=6$ steps).

When it comes to LSTM models (Fig.~\ref{subfig:how_much_data_is_needed_c} and Fig.~\ref{subfig:how_much_data_is_needed_d}), two main observations can be made. First, LSTMs using real data can accurately predict traffic and perform very well even when using a limited training dataset (e.g., $K=10$). This is because the LSTM solution explicitly focuses on finding patterns in time series data, which is something that CNNs learn implicitly. Second, the utility of synthetic datasets becomes more limited in LSTMs than for CNNs, thus its generation and adoption must be considered carefully in this case. For $K=10$ APs, the largest synthetic datasets ($|\mathcal{D}_S| = 60$ days) lead to worse performance than the shorter ones ($|\mathcal{D}_S| = \{15,30\}$ days), which is counterintuitive and uncorrelated with what we observed for the CNN models. The situation is somewhat reversed when using a higher number of APs ($K=\{20, 50\}$) for training the LSTMs, which allows for more diversity in the data. Still, it is unclear the benefit of using increasingly larger synthetic datasets for training LSTMs, provided that keeping the datasets small ($|\mathcal{D}_S| = 15$ days) can in some cases be the best among all the synthetic model solutions.

\subsection{When can synthetic models be used?}
\label{sec:exp3}

Next, we delve into the situations in which synthetically generated data are suitable for training ML models or, on the contrary, do not lead to appropriate results. For that, we first focus on the target test set of APs ($\Phi_{test}$) where the synthetically trained models are put into practice. To do this exercise, we focus on two possible options:
\begin{enumerate}
    \item $\Phi_{test} = \Phi_{train}$: The target test set belongs to the same set of APs ($\Phi_{train}$) whose data is used to train the models. This case is useful to represent the cases where personalized ML models---i.e., models trained using past measurements from the same target APs---are adopted.
    \item $\Phi_{test} \neq \Phi_{train}$: A different set of APs than the ones used to train the models ($\Phi_{train}$) is considered to evaluate predictive performance. This case illustrates well the situations in which pre-trained ML models are applied to different (unseen) deployments.
\end{enumerate}

\begin{figure*}[t!]
    \centering
    \begin{subfigure}{0.5\columnwidth}
        \centering
        \includegraphics[width=\linewidth]{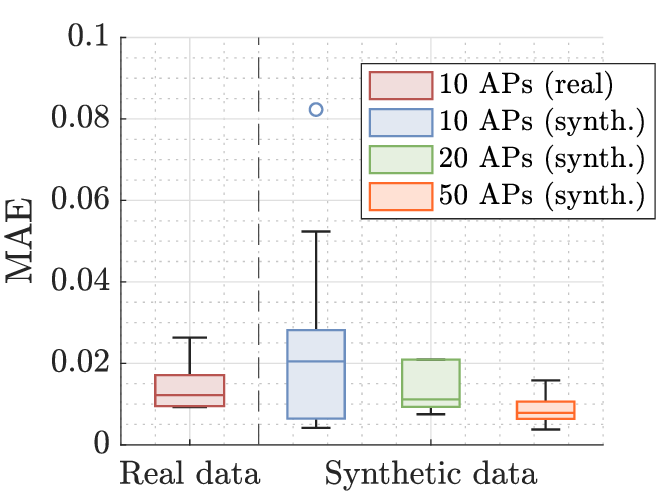}
        \caption{CNN, $\Phi_{test} = \Phi_{train}$, $s = 1$}
        \label{subfig:when_a}
    \end{subfigure}
    \begin{subfigure}{0.5\columnwidth}
        \centering
        \includegraphics[width=\linewidth]{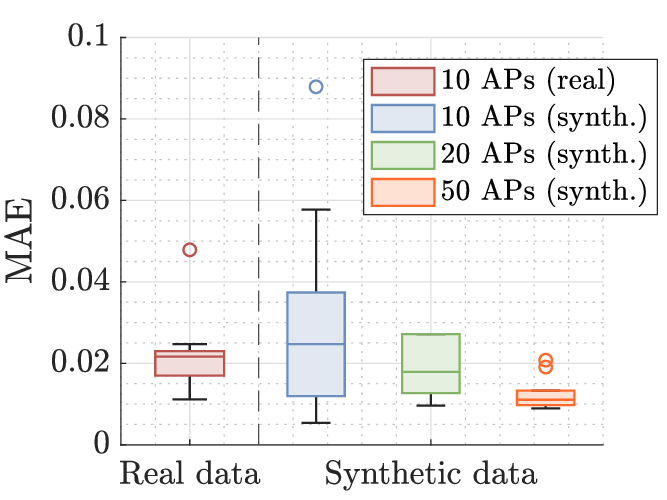}
        \caption{CNN, $\Phi_{test} = \Phi_{train}$, $s=6$}
        \label{subfig:when_b}
    \end{subfigure}
    \begin{subfigure}{0.5\columnwidth}
        \centering
        \includegraphics[width=\linewidth]{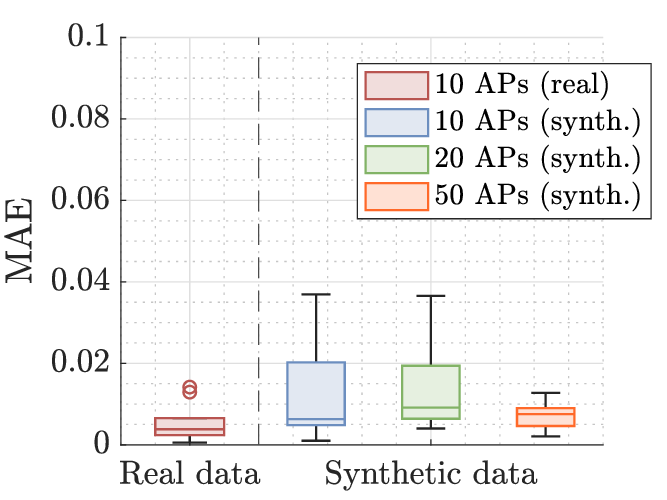}
        \caption{LSTM, $\Phi_{test} = \Phi_{train}$, $s=1$}
        \label{subfig:when_c}
    \end{subfigure}
    \begin{subfigure}{0.5\columnwidth}
        \centering
        \includegraphics[width=\linewidth]{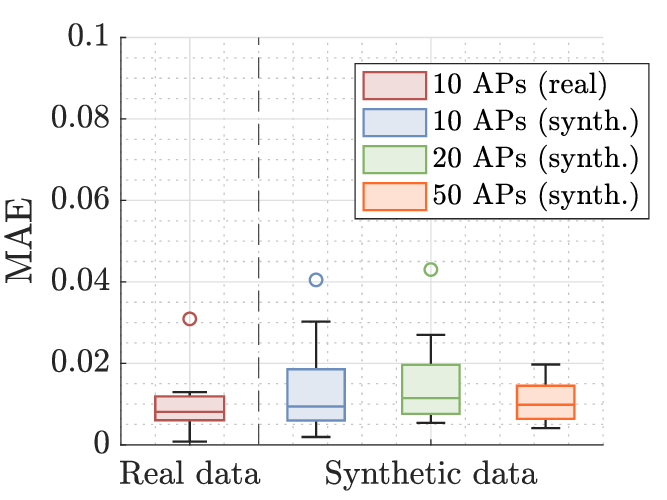}
        \caption{LSTM, $\Phi_{test} = \Phi_{train}$, $s=6$}
        \label{subfig:when_d}
    \end{subfigure}
    \begin{subfigure}{0.5\columnwidth}
        \centering
        \includegraphics[width=\linewidth]{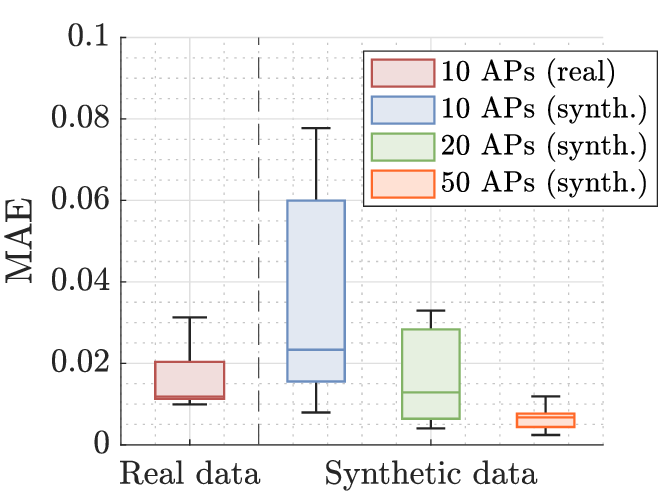}
        \caption{CNN, $\Phi_{test} \neq \Phi_{train}$, $s=1$}
        \label{subfig:when_e}
    \end{subfigure}
    \begin{subfigure}{0.5\columnwidth}
        \centering
        \includegraphics[width=\linewidth]{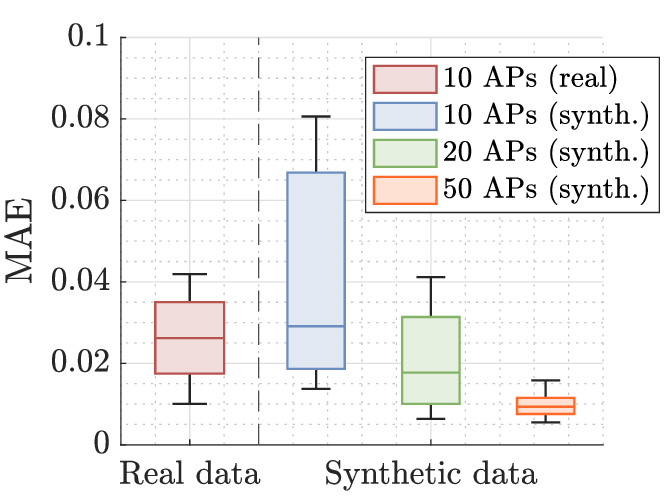}
        \caption{CNN, $\Phi_{test} \neq \Phi_{train}$, $s=6$}
        \label{subfig:when_f}
    \end{subfigure}
    \begin{subfigure}{0.5\columnwidth}
        \centering
        \includegraphics[width=\linewidth]{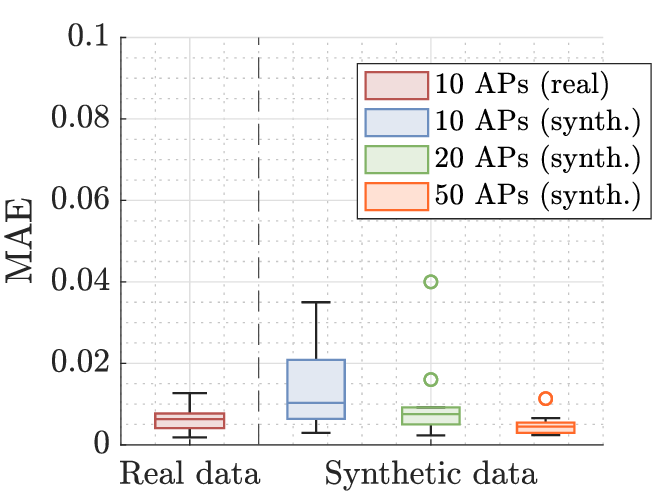}
        \caption{LSTM, $\Phi_{test} \neq \Phi_{train}$, $s=1$}
        \label{subfig:when_g}
    \end{subfigure}
    \begin{subfigure}{0.5\columnwidth}
        \centering
        \includegraphics[width=\linewidth]{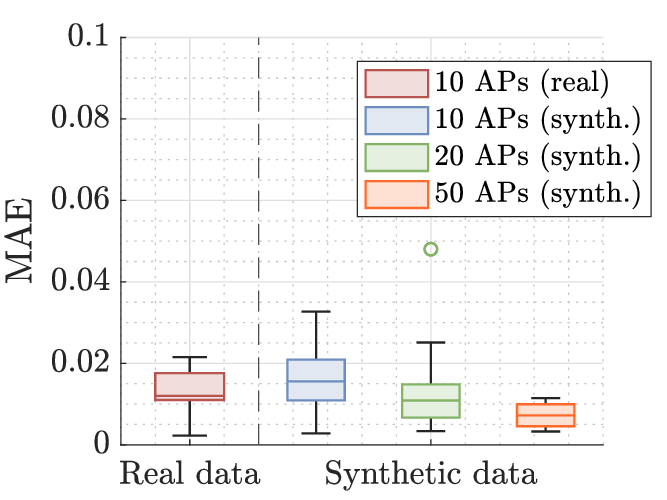}
        \caption{LSTM, $\Phi_{test} \neq \Phi_{train}$, $s=6$}
        \label{subfig:when_h}
    \end{subfigure}%
    \caption{\textit{When can synthetic-data-based models be used?} Predictive performance (MAE) achieved by each model ($m\in\{\texttt{CNN}, \texttt{LSTM}\}$) at different prediction horizons ($s \in\{1, 6\}$ steps) on different test sets ($\Phi_{test} = \Phi_{train}$ and $\Phi_{test} \neq \Phi_{train}$).}
    \label{fig:when_can_data_be_used}
\end{figure*}

Figure~\ref{fig:when_can_data_be_used} shows, again through boxplots, the MAE achieved in each setup ($m\in\{\texttt{CNN}, \texttt{LSTM}\}$, $s \in\{1, 6\}$ steps), $K\in\{10,20,50\}$). The main observations from Fig.~\ref{fig:when_can_data_be_used} are as follows.

\begin{itemize}
    \item $\Phi_{test} = \Phi_{train}$: In this case, a higher degree of specialization is achieved by the different ML models, provided that they are trained on the past data from APs where they are later put into practice. In order to perform similar to real models, synthetic models need to be trained with big enough synthetic datasets. Using data from $K=20$ synthetic APs is enough for CNN models (Fig.~\ref{subfig:when_a} and Fig.~\ref{subfig:when_b}), but $K=50$ is required in the LSTM cases (Fig.~\ref{subfig:when_c} and Fig.~\ref{subfig:when_d}). 
    \item $\Phi_{test} \neq \Phi_{train}$: In this case, where high generalization capacity is required, the need for potentially larger synthetic dataset becomes much more evident, which is palpable for $K=10$ in both the CNN (Fig.~\ref{subfig:when_e} and Fig.~\ref{subfig:when_f}) and the LSTM (Fig.~\ref{subfig:when_g} and Fig.~\ref{subfig:when_h}). In turn, when having access to plenty of synthetic data ($K=50$), the synthetic ML models significantly outperform the real ones, which is a key achievement.
\end{itemize}


To provide more insight into the feasibility of using synthetic data for training ML models, we show the false positives (i.e., the trained ML model predicts $\texttt{NO-LOAD}$ in a $\texttt{LOAD}$ situation) and false negatives (i.e., the trained ML model predicts $\texttt{LOAD}$ in a $\texttt{NO-LOAD}$ situation) achieved for different synthetic datasets. We refer to $\texttt{LOAD}$ and $\texttt{NO-LOAD}$ situations when, for a given time step, the AP load is above or below a threshold $\gamma$, respectively. The accurate prediction by an ML model of such a binary condition is important for triggering optimization actions, such as turning on/off an AP to save energy. Therefore, we consider a \texttt{LOAD} situation if the load ($l$) at a given time step in the future is $l \geq \gamma$. Likewise, a \texttt{NO-LOAD} situation is given by $l < \gamma$. Accordingly, we look at the cases where the predicted load, $l'$, leads to a wrong anticipation of \texttt{LOAD} (false positive) and \texttt{NO-LOAD} (false negative) situations.\footnote{Note that, for the energy-saving use case, false positives (the AP is switched off when it should not) might lead to a higher impact on the user than false negatives (the AP is not switched off when it could). Hence, mitigating false positives from ML predictions can be critical to ensuring a proper end-user experience.}

Regarding the results reported in Tables~\ref{table:false_positives} and \ref{table:false_negatives}, there are two basic premises that are valid in real models:
\begin{enumerate}
    \item A lower error (less false positives/negatives) is achieved as the problem becomes less challenging (bigger $\gamma$).
    \item A higher error (more false positives/negatives) is achieved as the number of prediction steps $s$ increases, which makes more difficult to predict the traffic accurately.
\end{enumerate}

These two key premises are also fulfilled by synthetic models, thus giving a first level of validity to the synthetic data. When it comes to more specific effects of the synthetic models in the different settings considered, we observe the following.
\begin{enumerate}
    \item \textbf{Synthetic dataset size ($K$):} The impact of the dataset size is not homogeneous across all the settings and must be carefully assessed. For instance, using larger synthetic datasets (a larger $K$) is counterproductive when the predictive task is challenging ($\gamma=0.01$ and $s=6$). In these cases, the number of false positives/negatives can increase up to the \num{22.59}/\num{92.36}\% for $K=50$. However, as the problem becomes more relaxed (higher $\gamma$, lower $s$), increasing $K$ leads to a lower error.  Still, there are cases where the impact that $K$ has on the accuracy of the models is unclear (see, for instance, the results achieved by the LSTM for $\gamma=0.05$ in Table~\ref{table:false_positives}), and this is due to the underlying complexity associated with the quality of the newly generated synthetic and their interplay with the rest of parameters. 
    \item \textbf{Generalization ($\Phi_{test} = \Phi_{train}$ vs. $\Phi_{test} \neq \Phi_{train}$):} 
    In line with the experiments from the previous sections, the error achieved for $\Phi_{test} = \Phi_{train}$ is typically lower than for $\Phi_{test} \neq \Phi_{train}$. Moreover, when the predictive task is challenging ($\gamma=0.01$, $s=6$, and $K=50$), the performance drop for $\Phi_{test} \neq \Phi_{train}$ is significantly high (e.g., for CNN, the false negative rate explodes to 92.36\%), thus throwing into question the usability of synthetic data in these cases. Nevertheless, there are exceptions in which synthetic models generalize well, which is particularly observed for the LSTM models (with a false negative rate of only 5.37\%, very close to the 4.48\% achieved by the LSTM trained on real data), demonstrating that synthetic models can achieve robust and generalizable performance.    
    \item \textbf{Model selection (CNN vs. LSTM):} Overall, the CNN models perform better in terms of false positives than false negatives (predict close to 0 load), making it more suitable for $\texttt{NO-LOAD}$ detection. When it comes to false positives, LSTM exhibits a better performance than CNN in many cases (false negatives rate often $<10$\%), thus making it more suitable for detecting higher load values. Moreover, we observe that the LSTM typically offers significantly better performance than CNN on $\Phi_{test} \neq \Phi_{train}$. If enough data is fed into the LSTM (e.g., $K=50$), it can offer reasonably good performance, even in the challenging cases described above ($\gamma=0.01$ and $s=6$). In that case, the LSTM's false positive rate is 12.42\%, significantly improving the CNN's 19.08\%.
\end{enumerate}

\begin{table*}[ht!]
\caption{False positives (\%) achieved in each setting when using real and synthetic data (in parenthesis, in \textcolor{cyan}{cyan}).}
\label{table:false_positives}
\resizebox{\textwidth}{!}{%
\begin{tabular}{@{}cc|ccc|ccc|ccc|ccc@{}}
\toprule
\multicolumn{2}{c}{\multirow{2}{*}{}} & \multicolumn{3}{c}{\textbf{CNN ($\Phi_{test} = \Phi_{train}$)}} & \multicolumn{3}{c}{\textbf{LSTM ($\Phi_{test} = \Phi_{train}$)}} & \multicolumn{3}{c}{\textbf{CNN ($\Phi_{test} \neq \Phi_{train}$)}} & \multicolumn{3}{c}{\textbf{LSTM ($\Phi_{test} \neq \Phi_{train}$)}} \\ \cmidrule(l){3-14} 
\multicolumn{2}{c}{} & $K = 10$ & $K = 20$ & $K = 50$ & $K = 10$ & $K = 20$ & $K = 50$ & $K = 10$ & $K = 20$ & $K = 50$ & $K = 10$ & $K = 20$ & $K = 50$ \\ \cmidrule(r){1-14}
\multirow{2}{*}{$\gamma = 0.01$} & $s = 1$ & 4.89 \textcolor{cyan}{(8.56)} & 3.40 \textcolor{cyan}{(16.09)} & 3.41 \textcolor{cyan}{(13.56)} & 5.60 \textcolor{cyan}{(9.71)} & 3.31 \textcolor{cyan}{(16.38)} & 3.76 \textcolor{cyan}{(14.90)} &  7.47 \textcolor{cyan}{(20.38)} & 2.66 \textcolor{cyan}{(7.82)} & 2.82 \textcolor{cyan}{(9.85)}  & 7.87 \textcolor{cyan}{(20.27)} & 3.55 \textcolor{cyan}{(9.73)} & 3.18 \textcolor{cyan}{(10.47)}\\
\cmidrule(l){2-14} 
 & $s = 6$ & 6.47 \textcolor{cyan}{(10.65)} & 7.46 \textcolor{cyan}{(21.40)} & 5.02 \textcolor{cyan}{(22.59)} & 8.25 \textcolor{cyan}{(11.00)} & 7.18 \textcolor{cyan}{(17.13)} & 7.30 \textcolor{cyan}{(16.74)} &  11.42 \textcolor{cyan}{(23.23)} & 5.07 \textcolor{cyan}{(9.86)} & 4.06 \textcolor{cyan}{(19.08)} & 11.97 \textcolor{cyan}{(20.60)} & 7.01 \textcolor{cyan}{(11.38)} & 6.47 \textcolor{cyan}{(12.42)}\\
 \cmidrule(l){2-14} 
\multirow{2}{*}{$\gamma = 0.05$} & $s = 1$ & 2.26 \textcolor{cyan}{(4.03)} & 2.00 \textcolor{cyan}{(7.81)} & 1.31 \textcolor{cyan}{(8.58)} & 3.09 \textcolor{cyan}{(4.82)} & 1.73 \textcolor{cyan}{(8.73)} & 1.57 \textcolor{cyan}{(6.82)} &  3.14 \textcolor{cyan}{(8.21)} & 1.23 \textcolor{cyan}{(4.15)} & 1.35 \textcolor{cyan}{(5.75)}  & 3.23 \textcolor{cyan}{(8.14)} & 1.34 \textcolor{cyan}{(4.35)} & 1.25 \textcolor{cyan}{(4.68)} \\
\cmidrule(l){2-14} 
 & $s = 6$ & 3.53 \textcolor{cyan}{(4.88)} & 4.58 \textcolor{cyan}{(8.31)} & 3.03 \textcolor{cyan}{(8.75)} & 4.58 \textcolor{cyan}{(5.83)} & 5.36 \textcolor{cyan}{(10.54)} & 3.98 \textcolor{cyan}{(9.27)} &  5.38 \textcolor{cyan}{(9.6)} & 3.37 \textcolor{cyan}{(4.99)} & 2.72 \textcolor{cyan}{(6.28)}  & 6.42 \textcolor{cyan}{(9.80)} & 4.25 \textcolor{cyan}{(7.09)} & 3.29 \textcolor{cyan}{(6.75)}\\
 \cmidrule(l){2-14} 
\multirow{2}{*}{$\gamma = 0.1$} & $s = 1$ & 1.29 \textcolor{cyan}{(2.66)} & 1.29 \textcolor{cyan}{(5.10)} & 0.75 \textcolor{cyan}{(5.59)} & 1.84 \textcolor{cyan}{(2.75)} & 1.13 \textcolor{cyan}{(5.30)} & 0.79 \textcolor{cyan}{(4.03)} &  2.67 \textcolor{cyan}{(5.79)} & 1.11 \textcolor{cyan}{(3.16)} & 0.58 \textcolor{cyan}{(3.33)}  & 2.29 \textcolor{cyan}{(5.46)} & 1.10 \textcolor{cyan}{(3.30)} & 0.64 \textcolor{cyan}{(2.73)} \\
\cmidrule(l){2-14} 
 & $s = 6$ & 2.43 \textcolor{cyan}{(3.46)} & 3.58 \textcolor{cyan}{(5.93)} & 2.22 \textcolor{cyan}{(6.22)} & 3.06 \textcolor{cyan}{(3.74)} & 4.41 \textcolor{cyan}{(7.08)} & 2.71 \textcolor{cyan}{(6.09)} &  4.68 \textcolor{cyan}{(6.7)} & 3.04 \textcolor{cyan}{(3.98)} & 1.77 \textcolor{cyan}{(4.03)}  & 5.15 \textcolor{cyan}{(7.25)} & 3.67 \textcolor{cyan}{(5.63)} & 2.07 \textcolor{cyan}{(4.26)} \\ \cmidrule(l){1-14} 
\end{tabular}
}
\end{table*}

\begin{table*}[ht!]
\caption{False negatives (\%) achieved in each setting when using real and synthetic data (in parenthesis, in \textcolor{cyan}{cyan}).}
\label{table:false_negatives}
\resizebox{\textwidth}{!}{%
\begin{tabular}{@{}cc|ccc|ccc|ccc|ccc@{}}
\toprule
\multicolumn{2}{c}{\multirow{2}{*}{}} & \multicolumn{3}{c}{\textbf{CNN ($\Phi_{test} = \Phi_{train}$)}} & \multicolumn{3}{c}{\textbf{LSTM ($\Phi_{test} = \Phi_{train}$)}} & \multicolumn{3}{c}{\textbf{CNN ($\Phi_{test} \neq \Phi_{train}$)}} & \multicolumn{3}{c}{\textbf{LSTM ($\Phi_{test} \neq \Phi_{train}$)}} \\ \cmidrule(l){3-14} 
\multicolumn{2}{c}{} & $K = 10$ & $K = 20$ & $K = 50$ & $K = 10$ & $K = 20$ & $K = 50$ & $K = 10$ & $K = 20$ & $K = 50$ & $K = 10$ & $K = 20$ & $K = 50$ \\ \cmidrule(r){1-14}
\multirow{2}{*}{$\gamma = 0.01$} & $s = 1$ & 30.03 \textcolor{cyan}{(25.40)}  & 19.70 \textcolor{cyan}{(50.26)} & 27.90 \textcolor{cyan}{(18.62)} & 2.30 \textcolor{cyan}{(4.25)} & 2.98 \textcolor{cyan}{(13.16)} & 3.03 \textcolor{cyan}{(5.87)} & 32.25 \textcolor{cyan}{(28.16)} & 15.45 \textcolor{cyan}{(31.81)} & 20.12 \textcolor{cyan}{(20.10)} & 4.69 \textcolor{cyan}{(7.37)} &  2.16 \textcolor{cyan}{(7.09)} & 2.13 \textcolor{cyan}{(4.57)} \\
\cmidrule(l){2-14} 
 & $s = 6$ & 31.48 \textcolor{cyan}{(38.47)} & 38.59 \textcolor{cyan}{(73.98)} & 38.73 \textcolor{cyan}{(91.49)} & 3.72 \textcolor{cyan}{(4.62)} & 5.61 \textcolor{cyan}{(13.72)} & 5.40 \textcolor{cyan}{(6.77)} & 34.37 \textcolor{cyan}{(38.57)} & 32.11 \textcolor{cyan}{(72.69)} & 35.17 \textcolor{cyan}{(92.36)} & 7.15 \textcolor{cyan}{(3.82)} & 4.50 \textcolor{cyan}{(8.46)} & 4.48 \textcolor{cyan}{(5.37)} \\
 \cmidrule(l){2-14} 
\multirow{2}{*}{$\gamma = 0.05$} & $s = 1$ & 5.79 \textcolor{cyan}{(8.80)} & 3.87 \textcolor{cyan}{(8.52)} & 11.82 \textcolor{cyan}{(0.89)} & 0.86 \textcolor{cyan}{(1.94)} & 1.37 \textcolor{cyan}{(4.48)} & 1.12 \textcolor{cyan}{(0.80)} & 10.78 \textcolor{cyan}{(13.25)} & 2.02 \textcolor{cyan}{(6.22)} & 9.90 \textcolor{cyan}{(0.92)} & 1.44 \textcolor{cyan}{(0.81)} & 0.84 \textcolor{cyan}{(1.14)} & 0.79 \textcolor{cyan}{(1.07)}\\
\cmidrule(l){2-14} 
 & $s = 6$ & 11.30 \textcolor{cyan}{(11.80)} & 10.04 \textcolor{cyan}{(12.02)} & 10.34 \textcolor{cyan}{(3.23)} & 2.28 \textcolor{cyan}{(2.99)} & 3.25 \textcolor{cyan}{(5.81)} & 2.79 \textcolor{cyan}{(1.59)} & 16.34 \textcolor{cyan}{(18.05)} & 6.85 \textcolor{cyan}{(8.72)} & 7.74 \textcolor{cyan}{(3.13)} & 3.69 \textcolor{cyan}{(1.99)} & 2.07 \textcolor{cyan}{(2.12)} & 2.18 \textcolor{cyan}{(1.60)}\\
 \cmidrule(l){2-14} 
\multirow{2}{*}{$\gamma = 0.1$} & $s = 1$ & 2.69 \textcolor{cyan}{(4.92)} & 2.28 \textcolor{cyan}{(6.85)} & 7.89 \textcolor{cyan}{(0.45)} & 0.49 \textcolor{cyan}{(1.14)} & 1.09 \textcolor{cyan}{(1.80)} & 0.80 \textcolor{cyan}{(0.32)} & 4.20 \textcolor{cyan}{(7.70)} & 1.06 \textcolor{cyan}{(4.00)} & 4.46 \textcolor{cyan}{(0.35)} & 0.71 \textcolor{cyan}{(0.50)} & 0.56 \textcolor{cyan}{(0.34)} & 0.54 \textcolor{cyan}{(0.38)} \\
\cmidrule(l){2-14} 
 & $s = 6$ & 5.49 \textcolor{cyan}{(6.67)} & 5.56 \textcolor{cyan}{(8.38)} & 5.10 \textcolor{cyan}{(1.41)} & 1.67 \textcolor{cyan}{(1.81)} & 2.57 \textcolor{cyan}{(2.81)} & 2.01 \textcolor{cyan}{(0.89)} & 8.24 \textcolor{cyan}{(9.74)} & 3.41 \textcolor{cyan}{(5.46)} & 3.20 \textcolor{cyan}{(1.12)} & 2.13 \textcolor{cyan}{(1.32)} & 1.40 \textcolor{cyan}{(1.08)} & 1.43 \textcolor{cyan}{(0.78)} \\ \cmidrule(l){1-14} 
\end{tabular}
}
\end{table*}

\subsection{Lessons learned}

Based on the analysis provided, the generalization capabilities of synthetic models are highly conditional and depend almost entirely on two factors: the ML model used (e.g., LSTM vs. CNN) and the difficulty of the task (e.g., $s=1$ vs. $s=6$). On one hand, synthetic ML models struggle with generalizing to unseen APs ($\Phi_{test} \neq \Phi_{train}$) but show a big potential when applied to the same devices from which data are extracted ($\Phi_{test} = \Phi_{train}$). In practice, this suggests that solutions based on synthetic data could be applied to improve deployments locally. At the same time, their usage should be carefully considered when it comes to new (unseen) devices, where mistaken predictions may lead to undesired effects (e.g., raising false alarms).

In general, having access to larger synthetic datasets can help improve performance and generalization. When enough synthetic data is used (e.g., $K=50$, $|\mathcal{D}_S^{(k)}| = 60$ days), a synthetic ML model can outperform models trained with limited real data (e.g., $K=10$, $|\mathcal{D}_R^{(k)}| = 15$ days). However, how much synthetic data must be used requires further analysis, as we have already seen that increasing $K$ does not always ensure better performance. The trade-off between how similar or different synthetic data must be from real data remains unclear, which is mostly because different APs exhibit uneven patterns (e.g., very low load vs. very high and varying load). In this regard, the effect that more and more diverse synthetic data adds to ML models (e.g., noise) is unknown in many cases.

When it comes to the ML model architecture, synthetic CNNs can align with real ones even if a relatively small amount of synthetic data is used (e.g., $K=20$), but synthetic LSTMs require more exhaustive data (e.g., $K=50$) to perform well. If this is met, then LSTMs are a much more powerful tool than CNNs, thus leading to much better performance for this particular problem.

\section{Conclusions}
\label{sec:conclusions}

As the demand for data-driven network intelligence continues to grow, synthetic data generation will play an increasingly vital role in advancing predictive analytics, autonomous network management, and next-generation communication systems. This paper proposes a novel approach to generate synthetic data for AI, which is applied and assessed for the Wi-Fi traffic prediction problem. Through a comprehensive statistical analysis, we highlight the main similarities and differences between the target real data and the synthetic data using our approach. In addition, we provide a broad set of experiments to assess the effectiveness of the synthetic data for training ML models. Our results show that the choice of model architecture is critical when using synthetic data. We demonstrate that while a CNN-based model fails to generalize, an LSTM trained on our synthetic data can achieve robust performance and generalization capabilities that are comparable to a model trained on real data, even in challenging scenarios. The demonstrated ability for specific architectures like LSTMs to leverage this synthetic data makes this approach particularly valuable for privacy-sensitive environments where data collection is constrained. It serves as a valuable tool for advancing wireless network analytics while preserving user privacy, and further research in this area has the potential to yield even more advancements in privacy-preserving machine learning for wireless communications. Future work includes novel techniques for validating synthetic data, the evaluation of synthetic data into different ML model architectures, and the study of the privacy-preserving capabilities of the synthetic data approach.

\section*{Acknowledgement}

This work has been co-funded by a grant from Banco Santander directed to doctoral students of the University of Málaga's doctoral program for the international mention through sectoral action 321 of the Comprehensive Teaching Plan. The work of F. Wilhelmi is supported by the CHIST-ERA Wireless AI 2022 call MLDR project (ANR-23-CHR4-0005), partially funded by AEI under project PCI2023-145958-2, by Wi-XR PID2021-123995NB-I00 and TRUE-Wi-Fi PID2024-155470NB-I00 (MCIU/AEI/FEDER,UE), and by MCIN/AEI under the Maria de Maeztu Units of Excellence Programme (CEX2021-001195-M). The work of A. Fernandez Duran and L. Galati Giordano is partially supported by CDTI project IDI-20250211 MINERGY

\bibliographystyle{IEEEtran}
\bibliography{bib}

@IEEEtranBSTCTL{IEEEexample:BSTcontrol,
CTLuse_forced_etal       = "yes",
CTLmax_names_forced_etal = "2",
CTLnames_show_etal       = "1" }

@book{emam2020chap,
  author={Khaled El Emam and Lucy Mosquera and Richard Hoptroff},
  title={{Practical Synthetic Data Generation: Balancing Privacy and the Broad Availability of Data}},
  publisher={O'Reilly Media},
  year={2020}
}

@article{song2021federated,
  author={He Song and Qi Zhang and Kai Yang},
  title={{Federated Learning for Privacy-Preserving AI in Edge Computing: A Survey}},
  journal={ACM Computing Surveys},
  volume={54},
  number={7},
  year={2021},
  pages={1-36}
}

@article{xu2018deep,
  author={Rui Xu and Jie Li and Huan Liu},
  title={{Deep Learning for Network Traffic Analysis: A Review}},
  journal={IEEE Communications Surveys \& Tutorials},
  volume={20},
  number={4},
  pages={3453-3473},
  year={2018}
}

@article{yang2023bias,
  author={Xiang Yang and Mingzhe Wang and Lijun Sun},
  title={{Bias in Machine Learning-Based Network Traffic Prediction: Challenges and Mitigation Strategies}},
  journal={IEEE Transactions on Network and Service Management},
  year={2023},
  volume={20},
  number={1},
  pages={128-141}
}

@article{goncalves2020generation,
  author={André Gonçalves and Luís Antunes and Pedro Bizarro},
  title={{Generation and Evaluation of Synthetic Data: Methods and Applications}},
  journal={Artificial Intelligence Review},
  year={2020},
  volume={53},
  pages={2857-2888}
}

@article{bega2019deepcog,
  author={Dario Bega and Marco Gramaglia and Xavier Costa-Pérez},
  title={{DeepCog: Cognitive Network Management in Sliced 5G Networks with Deep Learning}},
  journal={IEEE Journal on Selected Areas in Communications},
  volume={37},
  number={10},
  pages={2372-2382},
  year={2019}
}

@article{giordani2020toward,
  author={Marco Giordani and Marco Mezzavilla and Michele Polese},
  title={{Toward 6G Networks: Use Cases and Technologies}},
  journal={IEEE Communications Magazine},
  volume={58},
  number={3},
  pages={55-61},
  year={2020}
}

@misc{ammara2025syn,
      title={{Synthetic Network Traffic Data Generation: A Comparative Study}}, 
      author={Dure Adan Ammara and Jianguo Ding and Kurt Tutschku},
      year={2025},
      eprint={2410.16326},
      archivePrefix={arXiv},
      primaryClass={cs.CR},
      url={https://arxiv.org/abs/2410.16326}
}

@article{naveed2021is,
title={{Is Synthetic The New Real? Performance Analysis of Time Series Generation Techniques with Focus on Network Load Forecasting}},
url={http://dx.doi.org/10.36227/techrxiv.17296235.v1},
DOI={10.36227/techrxiv.17296235.v1},
publisher={Institute of Electrical and Electronics Engineers (IEEE)},
author={Naveed, Muhammad Haris and Hashmi, Umair and Tajved, Nayab and Sultan, Neha and Imran, Ali},
year={2021}
}

@misc{zhao2025does,
      title={{Does Training with Synthetic Data Truly Protect Privacy?}}, 
      author={Yunpeng Zhao and Jie Zhang},
      year={2025},
      eprint={2502.12976},
      archivePrefix={arXiv},
      primaryClass={cs.CR},
      url={https://arxiv.org/abs/2502.12976}
}

@inproceedings{basu1996time,
  title={{Time series models for internet traffic}},
  author={Basu, Sabyasachi and Mukherjee, Amarnath and Klivansky, Steve},
  booktitle={Proceedings of IEEE INFOCOM'96. Conference on Computer Communications},
  volume={2},
  pages={611--620},
  year={1996},
  organization={IEEE}
}

@inproceedings{rotem2022transfer,
  title={{Transfer learning for time series classification using synthetic data generation}},
  author={Rotem, Yarden and Shimoni, Nathaniel and Rokach, Lior and Shapira, Bracha},
  booktitle={International Symposium on Cyber Security, Cryptology, and Machine Learning},
  pages={232--246},
  year={2022},
  organization={Springer}
}

@article{ren2019inf,
author = {Ren, Jingjing and Dubois, Daniel and Choffnes, David and Mandalari, Anna and Kolcun, Roman and Haddadi, Hamed},
year = {2019},
month = {10},
pages = {267-279},
title = {{Information Exposure From Consumer IoT Devices: A Multidimensional, Network-Informed Measurement Approach}},
journal = {IMC '19: Proceedings of the Internet Measurement Conference}
}

@misc{ostinato2010,
  title     = {{Ostinato: Traffic generator for network engineers}},
  year      = {2010},
  note      = {[Online]. Available: \url{https://ostinato.org}}
}

@misc{rude2000,
  title     = {{Rude \& Crude}},
  year      = {2000},
  note      = {[Online]. Available: \url{https://rude.sourceforge.net}}
}

@misc{seagull2006,
  title     = {{Seagull: An Open Source Multi-Protocol Traffic Generator}},
  year      = {2006},
  note      = {[Online]. Available: \url{https://gull.sourceforge.net}}
}

@misc{ns3_2011,
  title     = {{Ns-3 Network Simulator}},
  year      = {2011},
  note      = {[Online]. Available: \url{https://www.nsnam.org/}}
}

@article{xu2020stan,
  author    = {S. Xu and others},
  title     = {{{STAN}: Synthetic Network Traffic Generation Using Autoregressive Neural Models}},
  year      = {2020},
  archivePrefix = {arXiv},
  eprint    = {2009.12740},
  primaryClass = {cs.NI},
}

@inproceedings{wang2020packetcgan,
  author    = {P. Wang and S. Li and F. Ye and Z. Wang and M. Zhang},
  title     = {{PacketCGAN: Exploratory Study of Class Imbalance for Encrypted Traffic Classification Using CGAN}},
  booktitle = {Proc. IEEE Int. Conf. Commun.},
  year      = {2020},
  pages     = {1-7},
}

@inproceedings{wilhelmi2023ai,
  title={{AI/ML-based Load Prediction in IEEE 802.11 Enterprise Networks}},
  author={Wilhelmi, Francesc and Salami, Dariush and Fontanesi, Gianluca and Galati-Giordano, Lorenzo and Kasslin, Mika},
  booktitle={IEEE International Conference on Machine Learning for Communication and Networking},
  year={2024}
}

@article{chen2021flag,
  title={{Flag: Flexible, accurate, and long-time user load prediction in large-scale WiFi system using deep RNN}},
  author={Chen, Wenxiong and Lyu, Feng and Wu, Fan and Yang, Peng and Ren, Ju},
  journal={IEEE Internet of Things Journal},
  volume={8},
  number={22},
  pages={16510--16521},
  year={2021},
  publisher={IEEE}
}

@ARTICLE{li2024IoT,
  author={Li, Ruoyu and Li, Qing and Zou, Qingsong and Zhao, Dan and Zeng, Xiangyi and Huang, Yucheng and Jiang, Yong and Lyu, Feng and Ormazabal, Gaston and Singh, Aman and Schulzrinne, Henning},
  journal={IEEE Transactions on Mobile Computing}, 
  title={{IoTGemini: Modeling IoT Network Behaviors for Synthetic Traffic Generation}}, 
  year={2024},
  volume={23},
  number={12},
  pages={13240-13257},
}

@article{figueira2022survey,
  title={{Survey on Synthetic Data Generation, Evaluation Methods and GANs}},
  author={Figueira, Alvaro and Bruno Vaz},
  journal={Mathematics},
  volume={10},
  number={15},
  pages={2733},
  year={2022}
}

@ARTICLE{chai2024gen,
  author={Chai, Haoye and Wang, Huandong and Li, Tong and Wang, Zhaocheng},
  journal={IEEE Network}, 
  title={Generative {AI}-Driven {D}igital {T}win for {M}obile {N}etworks}, 
  year={2024},
  volume={38},
  number={5},
  pages={84-92},
  doi={10.1109/MNET.2024.3420702}}

@ARTICLE{zhang2018gen,
  author={Zhang, Chi and Kuppannagari, Sanmukh R. and Kannan, Rajgopal and Prasanna, Viktor K.},
  journal={2018 {IEEE} International Conference on Communications, Control, and Computing Technologies for Smart Grids (SmartGridComm)}, 
  title={Generative Adversarial Network for Synthetic Time Series Data Generation in Smart Grids}, 
  year={2018},
  volume={},
  number={},
  pages={1-6},
  doi={10.1109/SmartGridComm.2018.8587464}}

@Article{subhajit2023syn,
AUTHOR = {Chatterjee, Subhajit and Byun, Yung-Cheol},
TITLE = {A Synthetic Data Generation Technique for Enhancement of Prediction Accuracy of Electric Vehicles Demand},
JOURNAL = {Sensors},
VOLUME = {23},
YEAR = {2023},
NUMBER = {2},
ARTICLE-NUMBER = {594}}

@article{ahmad2020machine,
  title={{Machine learning meets communication networks: Current trends and future challenges}},
  author={Ahmad, Ijaz and Shahabuddin, Shariar and Malik, Hassan and Harjula, Erkki and Lepp{\"a}nen, Teemu and Loven, Lauri and Anttonen, Antti and Sodhro, Ali Hassan and Alam, Muhammad Mahtab and Juntti, Markku and others},
  journal={IEEE Access},
  volume={8},
  pages={223418--223460},
  year={2020},
  publisher={IEEE}
}

@article{liu2019machine,
  title={{When machine learning meets big data: A wireless communication perspective}},
  author={Liu, Yuanwei and Bi, Suzhi and Shi, Zhiyuan and Hanzo, Lajos},
  journal={IEEE Vehicular Technology Magazine},
  volume={15},
  number={1},
  pages={63--72},
  year={2019},
  publisher={IEEE}
}

@misc{dimyati2021time,
  title={{Time-series generative adversarial networks for telecommunications data augmentation}},
  author={Dimyati, Hamid},
  year={2021}
}

@article{pandey20245gt,
  title={{5GT-GAN-NET: Internet Traffic Data Forecasting With Supervised Loss Based Synthetic Data Over 5G}},
  author={Pandey, Chandrasen and Tiwari, Vaibhav and Rodrigues, Joel JPC and Roy, Diptendu Sinha},
  journal={IEEE Transactions on Mobile Computing},
  year={2024},
  publisher={IEEE}
}

@article{wilhelmi2022federated,
  title={{Federated spatial reuse optimization in next-generation decentralized IEEE 802.11 WLANs}},
  author={Wilhelmi, Francesc and Hribar, Jernej and Yilmaz, Selim F and Ozfatura, Emre and Ozfatura, Kerem and Yildiz, Ozlem and G{\"u}nd{\"u}z, Deniz and Chen, Hao and Ye, Xiaoying and You, Lizhao and others},
  journal={ITU-T Journal on Emerging Technologies},
  volume={3},
  number={2},
  pages={117-133},
  year={2022}
}

@article{pulido2025digital,
  author={Pulido, José and Fortes, Sergio and Barco, Raquel},
  booktitle={2025 IEEE 21st International Conference on Factory Communication Systems (WFCS)}, 
  title={{Digital Twin-Based in Next-Generation Wi-Fi Networks: Survey and Future Challenges}}, 
  year={2025},
  volume={},
  number={},
  pages={1-8},
  doi={10.1109/WFCS63373.2025.11077636}}

@inproceedings{patki2016synthetic,
  title={{The synthetic data vault}},
  author={Patki, Neha and Wedge, Roy and Veeramachaneni, Kalyan},
  booktitle={2016 IEEE international conference on data science and advanced analytics (DSAA)},
  pages={399--410},
  year={2016},
  organization={IEEE}
}

@misc{lu2021mach,
      title={{Machine Learning for Synthetic Data Generation: A Review}}, 
      author={Yingzhou Lu and Lulu Chen and Yuanyuan Zhang and Minjie Shen and Huazheng Wang and Xiao Wang and Capucine van Rechem and Tianfan Fu and Wenqi Wei},
      year={2025},
      eprint={2302.04062},
      archivePrefix={arXiv},
      primaryClass={cs.LG},
      url={https://arxiv.org/abs/2302.04062}, 
}

@inproceedings{omnetpp2001,
  title={{OMNeT++: Objective Modular Network Testbed in C++}},
  author={Varga, András},
  booktitle={Proceedings of the 15th European Simulation Multiconference (ESM)},
  year={2001}
}

@inproceedings{wilhelmi2021komondor,
  title={{Komondor: A Wireless Network Simulator for Next-Generation High-Density WLANs}},
  author={Wilhelmi, Francesc and Barrachina-Muñoz, Sònia and Gringoli, Francesco and Fiore, Marco},
  booktitle={IEEE Conference on Computer Communications Workshops (INFOCOM WKSHPS)},
  year={2021},
  pages={1--6},
  publisher={IEEE}
}

@inproceedings{krasic2022telecom,
  title={{Telecom fraud detection with machine learning on imbalanced dataset}},
  author={Krasi{\'c}, Ivan and {\v{C}}elar, Stipe},
  booktitle={2022 International Conference on Software, Telecommunications and Computer Networks (SoftCOM)},
  pages={1--6},
  year={2022},
  organization={IEEE}
}

@article{wilhelmi2024s,
  title={{``It's Your Turn": A Novel Channel Contention Mechanism for Improving Wi-Fi's Reliability}},
  author={Wilhelmi, Francesc and Galati-Giordano, Lorenzo and Fontanesi, Gianluca},
  journal={arXiv preprint arXiv:2410.07874},
  year={2024}
}

@article{fauzi2022mobile,
  title={{Mobile network coverage prediction based on supervised machine learning algorithms},
  author={Fauzi, Mohd Fazuwan Ahmad and Nordin, Rosdiadee and Abdullah, Nor Fadzilah and Alobaidy, Haider AH}},
  journal={IEEE Access},
  volume={10},
  pages={55782--55793},
  year={2022},
  publisher={IEEE}
}

@inproceedings{hussain2010capacity,
  title={{Capacity planning of network redesign—A case study}},
  author={Hussain, Tahani H and Habib, Sami J},
  booktitle={Proceedings of the 2010 International Symposium on Performance Evaluation of Computer and Telecommunication Systems (SPECTS'10)},
  pages={52--57},
  year={2010},
  organization={IEEE}
}

@article{hoydis2020toward,
  title={{Toward a 6G AI-native air interface}},
  author={Hoydis, Jakob and Aoudia, Fay{\c{c}}al Ait and Valcarce, Alvaro and Viswanathan, Harish},
  journal={arXiv preprint arXiv:2012.08285},
  year={2020}
}

@inproceedings{zhu2021network,
  title={{Network planning with deep reinforcement learning}},
  author={Zhu, Hang and Gupta, Varun and Ahuja, Satyajeet Singh and Tian, Yuandong and Zhang, Ying and Jin, Xin},
  booktitle={Proceedings of the 2021 ACM SIGCOMM 2021 Conference},
  pages={258--271},
  year={2021}
}

@article{bellalta2024towards,
  title={{Towards an AI/ML-defined Radio for Wi-Fi: Overview, Challenges, and Roadmap}},
  author={Bellalta, Boris and Kosek-Szott, Katarzyna and Szott, Szymon and Wilhelmi, Francesc},
  journal={arXiv preprint arXiv:2405.12675},
  year={2024}
}

@article{wilhelmi2021usage,
  title={{Usage of network simulators in machine-learning-assisted 5G/6G networks}},
  author={Wilhelmi, Francesc and Carrascosa, Marc and Cano, Cristina and Jonsson, Anders and Ram, Vishnu and Bellalta, Boris},
  journal={IEEE Wireless Communications},
  volume={28},
  number={1},
  pages={160--166},
  year={2021},
  publisher={IEEE}
}

@inproceedings{gawlowicz2019ns,
  title={{Ns-3 meets openai gym: The playground for machine learning in networking research}},
  author={Gaw{\l}owicz, Piotr and Zubow, Anatolij},
  booktitle={Proceedings of the 22nd International ACM Conference on Modeling, Analysis and Simulation of Wireless and Mobile Systems},
  pages={113--120},
  year={2019}
}

@article{feuerriegel2024generative,
  title={{Generative AI}},
  author={Feuerriegel, Stefan and Hartmann, Jochen and Janiesch, Christian and Zschech, Patrick},
  journal={Business \& Information Systems Engineering},
  volume={66},
  number={1},
  pages={111--126},
  year={2024},
  publisher={Springer}
}

@article{fernandes2019comprehensive,
  title={{A comprehensive survey on network anomaly detection}},
  author={Fernandes Jr, Gilberto and Rodrigues, Joel JPC and Carvalho, Luiz Fernando and Al-Muhtadi, Jalal F and Proen{\c{c}}a Jr, Mario Lemes},
  journal={Telecommunication Systems},
  volume={70},
  number={3},
  pages={447--489},
  year={2019},
  publisher={Springer}
}

\end{document}